\documentclass{egpubl}
\usepackage{eg2017}

\ConferenceSubmission

 \electronicVersion %

\ifpdf \usepackage[pdftex]{graphicx} \pdfcompresslevel=9
\else \usepackage[dvips]{graphicx} \fi

\PrintedOrElectronic

\usepackage{t1enc,dfadobe}

\usepackage{egweblnk}
\usepackage{cite}

\usepackage{url}
\usepackage[table]{xcolor}
\usepackage{multirow}
\usepackage{bigstrut}
\usepackage{transparent}
\definecolor{rev}{rgb}{0,0,0}
\newcommand{\specialcell}[2][c]{%
  \begin{tabular}[#1]{@{}c@{}}#2\end{tabular}}

\title{Performance-Based Biped Control using a Consumer Depth Camera}

\author[Y. Lee \& T. Kwon]
       {Yoonsang Lee$^{1}$
        and Taesoo Kwon$^{2}$
        \\
         $^1$Kwangwoon University \ \ \ \
         $^2$Hanyang University
       }

\begin{document}

 \teaser{
  \includegraphics[width=\linewidth]{eg_new}
  \centering
   \caption{New EG Logo}
 \label{fig:teaser}
 }

 \teaser{
	\centering
    \reflectbox{\includegraphics[trim=880 170 150 170, clip, width=.13\linewidth]{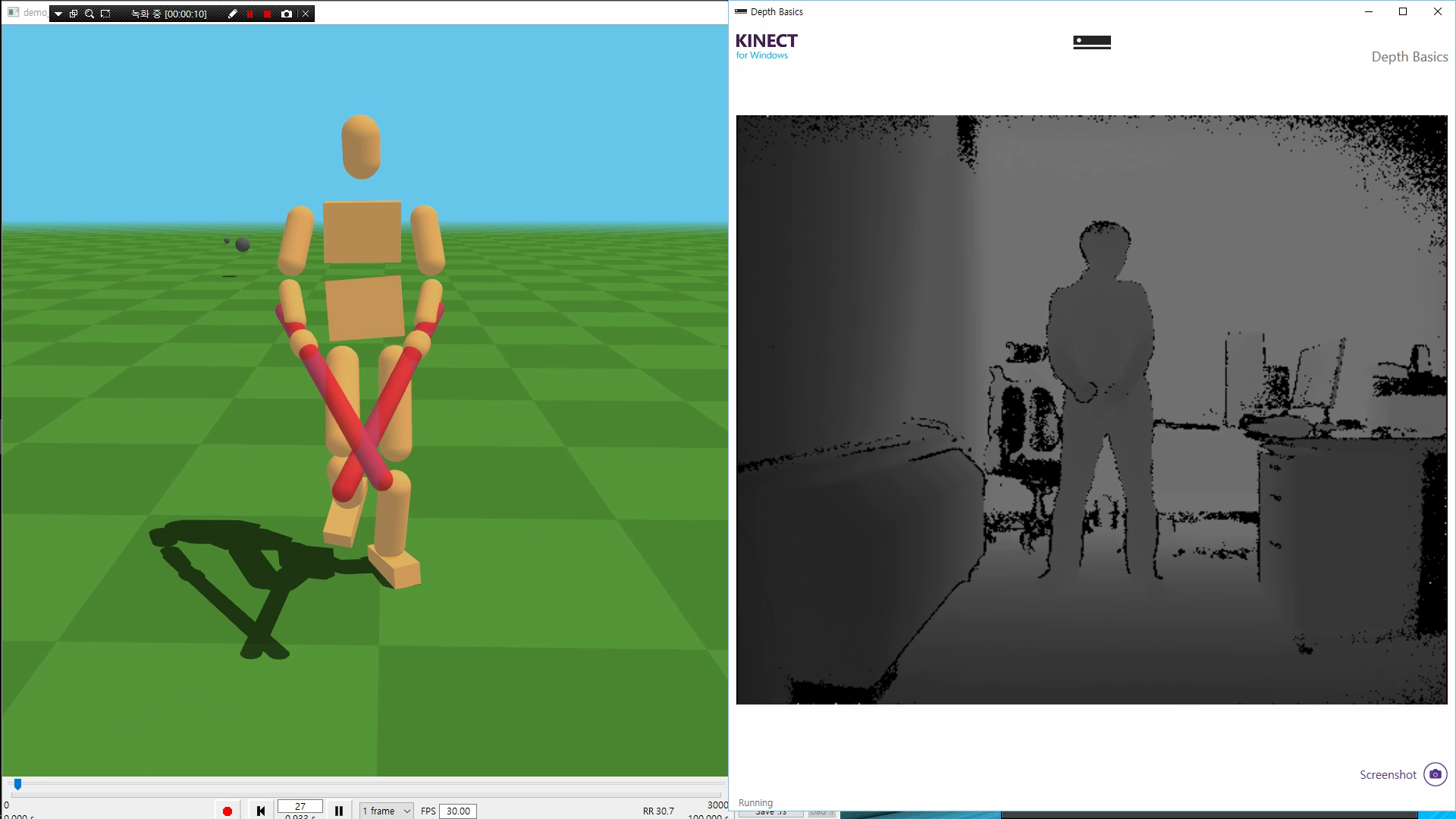}}
    \reflectbox{\includegraphics[trim=880 170 150 170, clip, width=.13\linewidth]{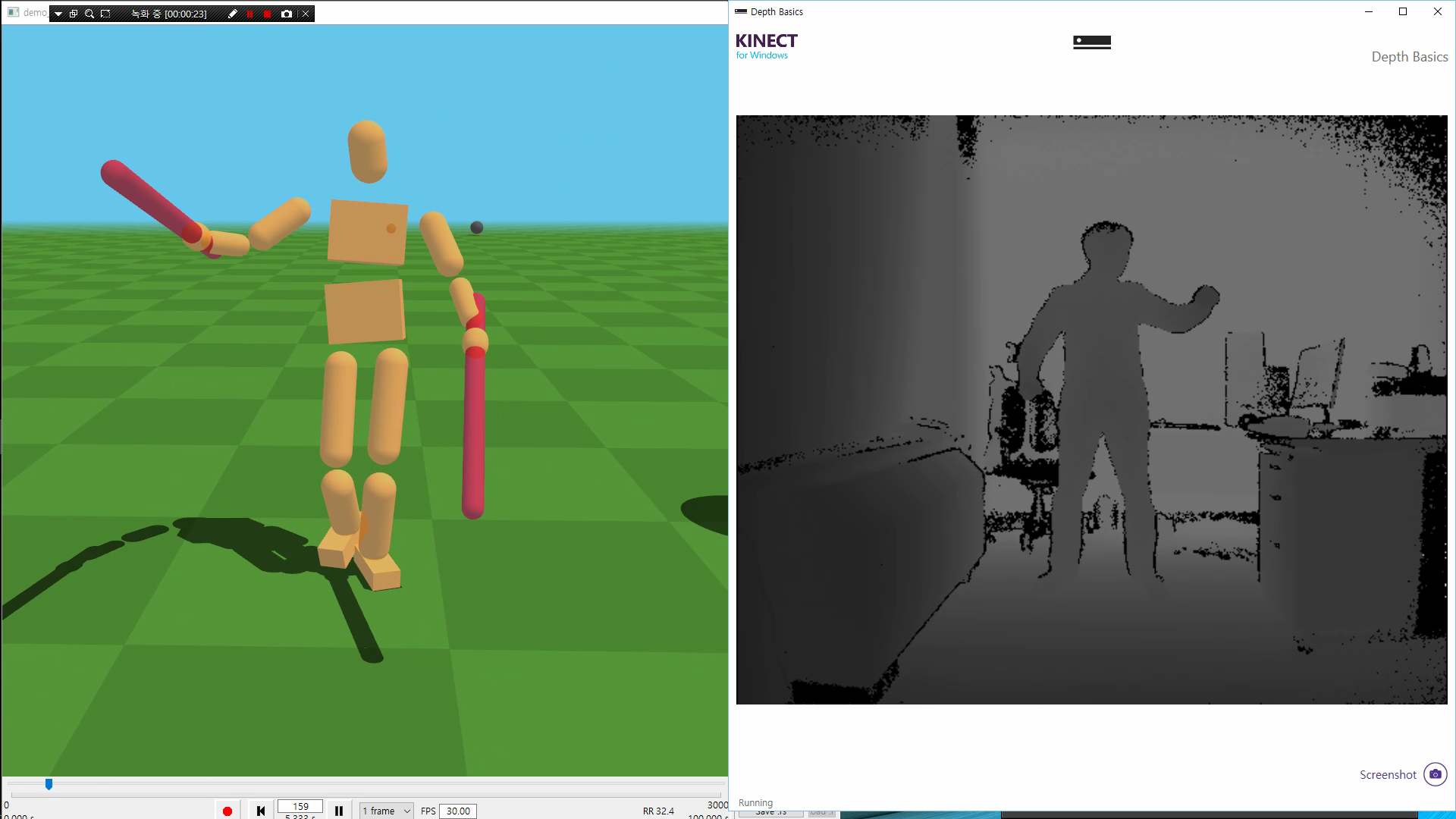}}
    \reflectbox{\includegraphics[trim=880 170 150 170, clip, width=.13\linewidth]{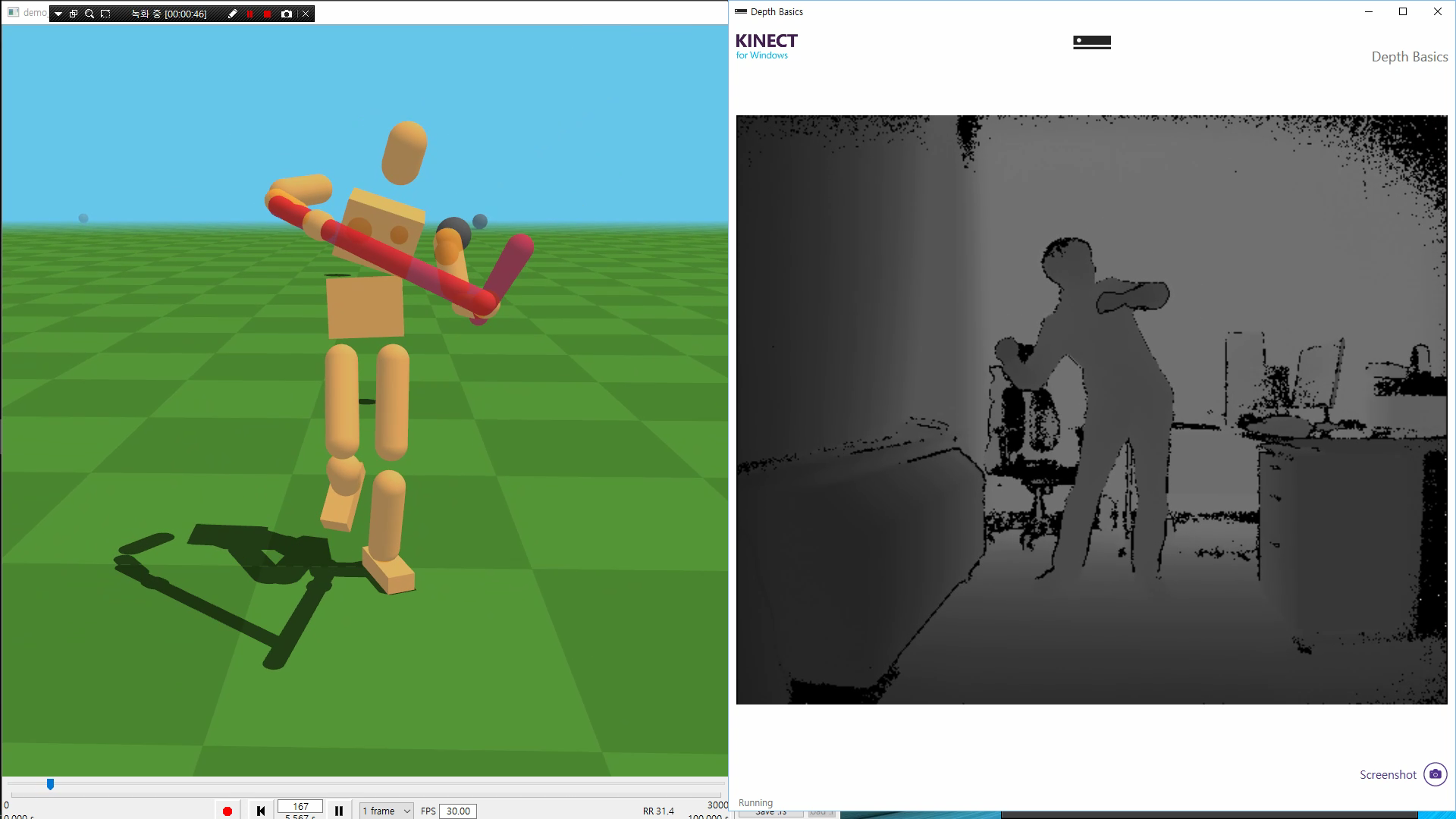}}
    \reflectbox{\includegraphics[trim=880 170 150 170, clip, width=.13\linewidth]{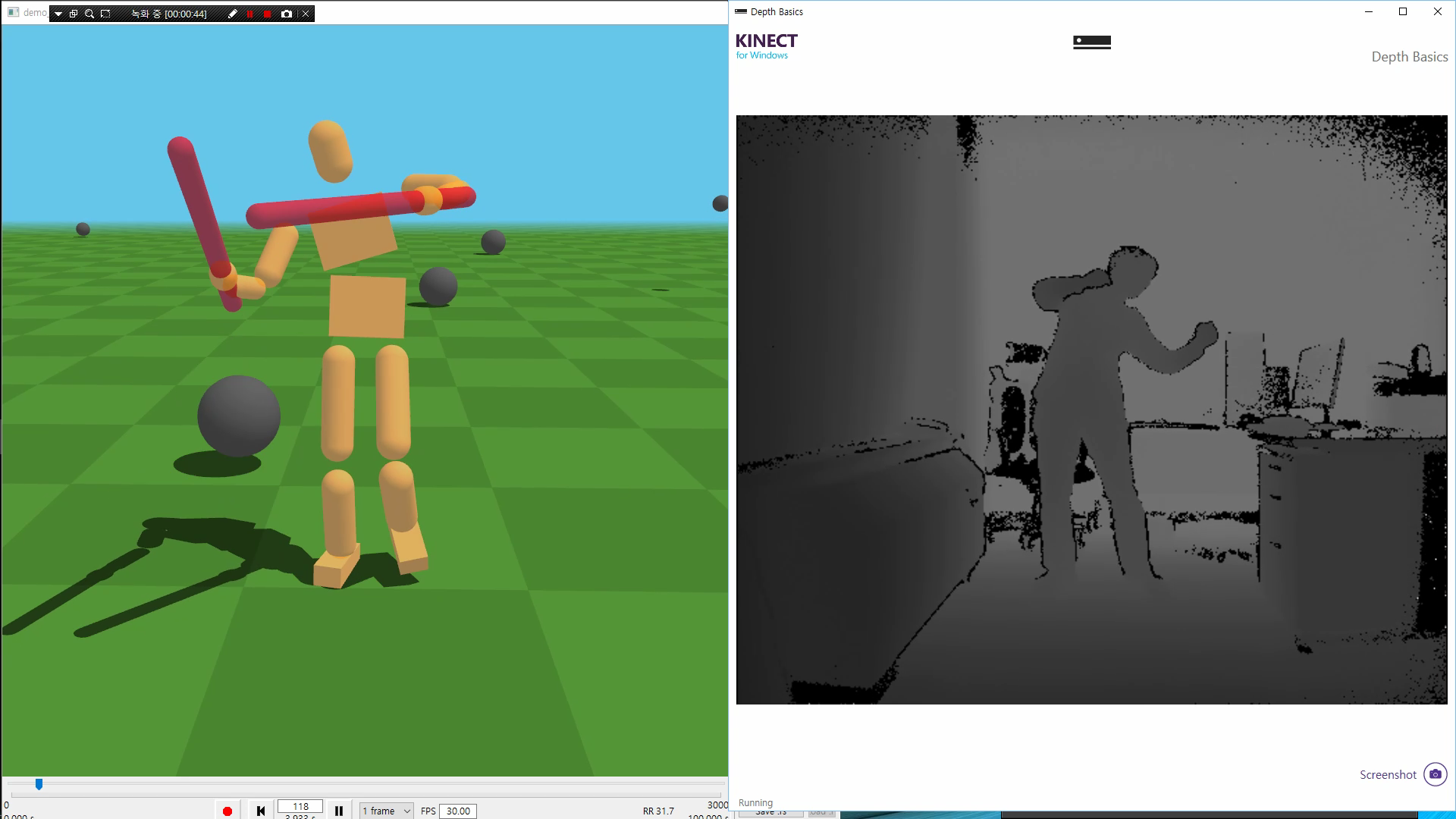}}
    \reflectbox{\includegraphics[trim=880 170 150 170, clip, width=.13\linewidth]{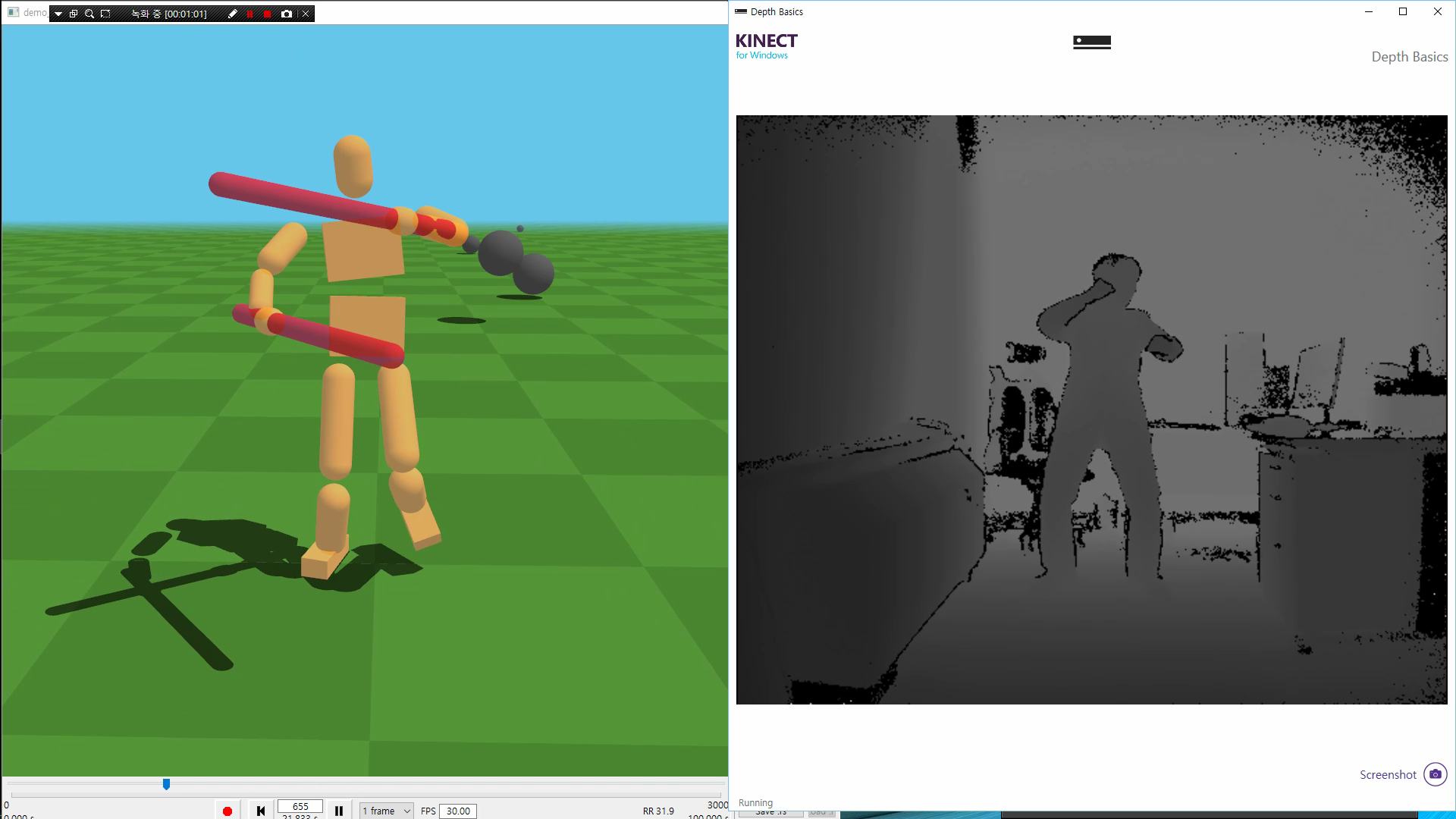}}
    \reflectbox{\includegraphics[trim=880 170 150 170, clip, width=.13\linewidth]{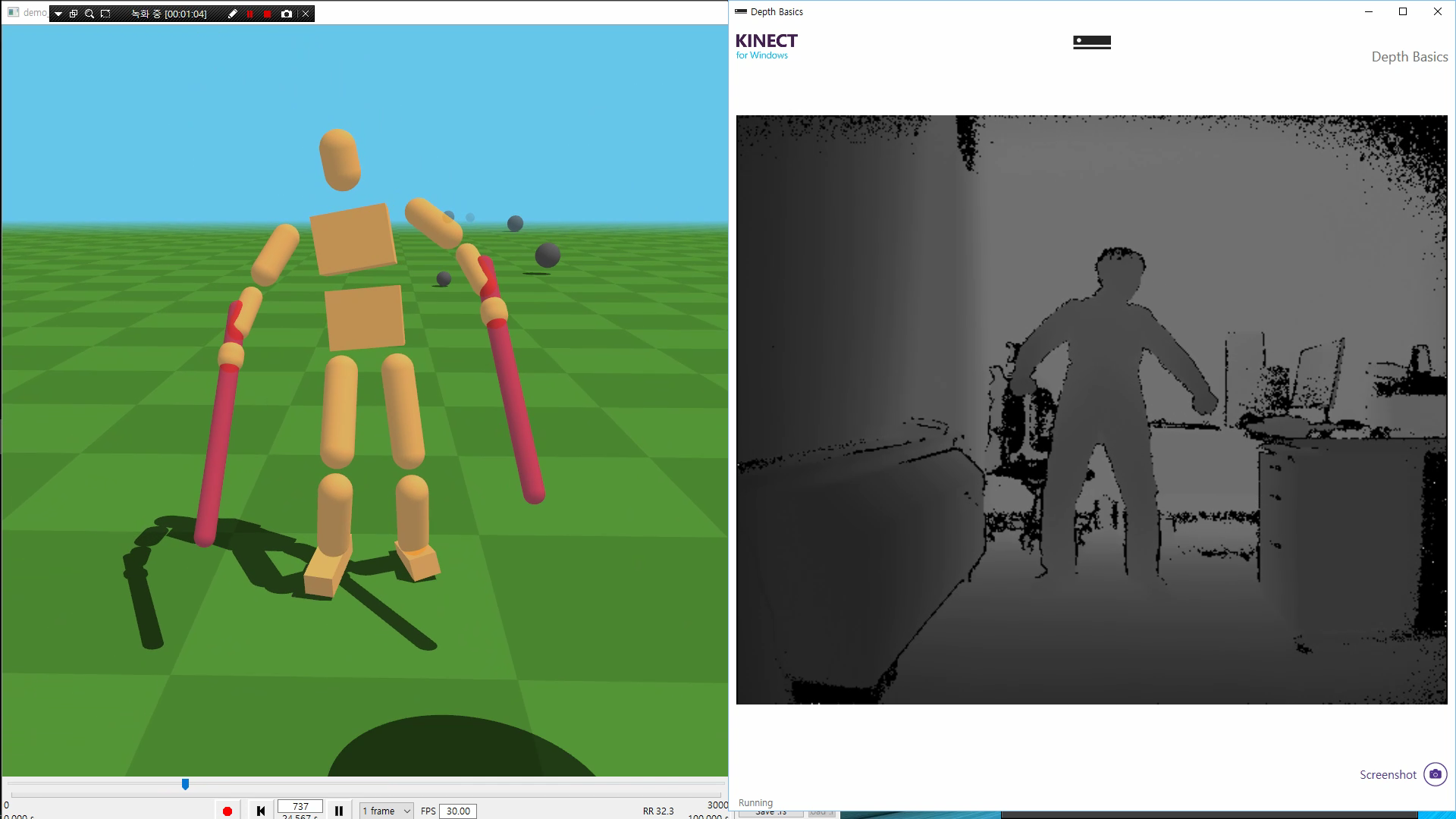}}
    \reflectbox{\includegraphics[trim=880 170 150 170, clip, width=.13\linewidth]{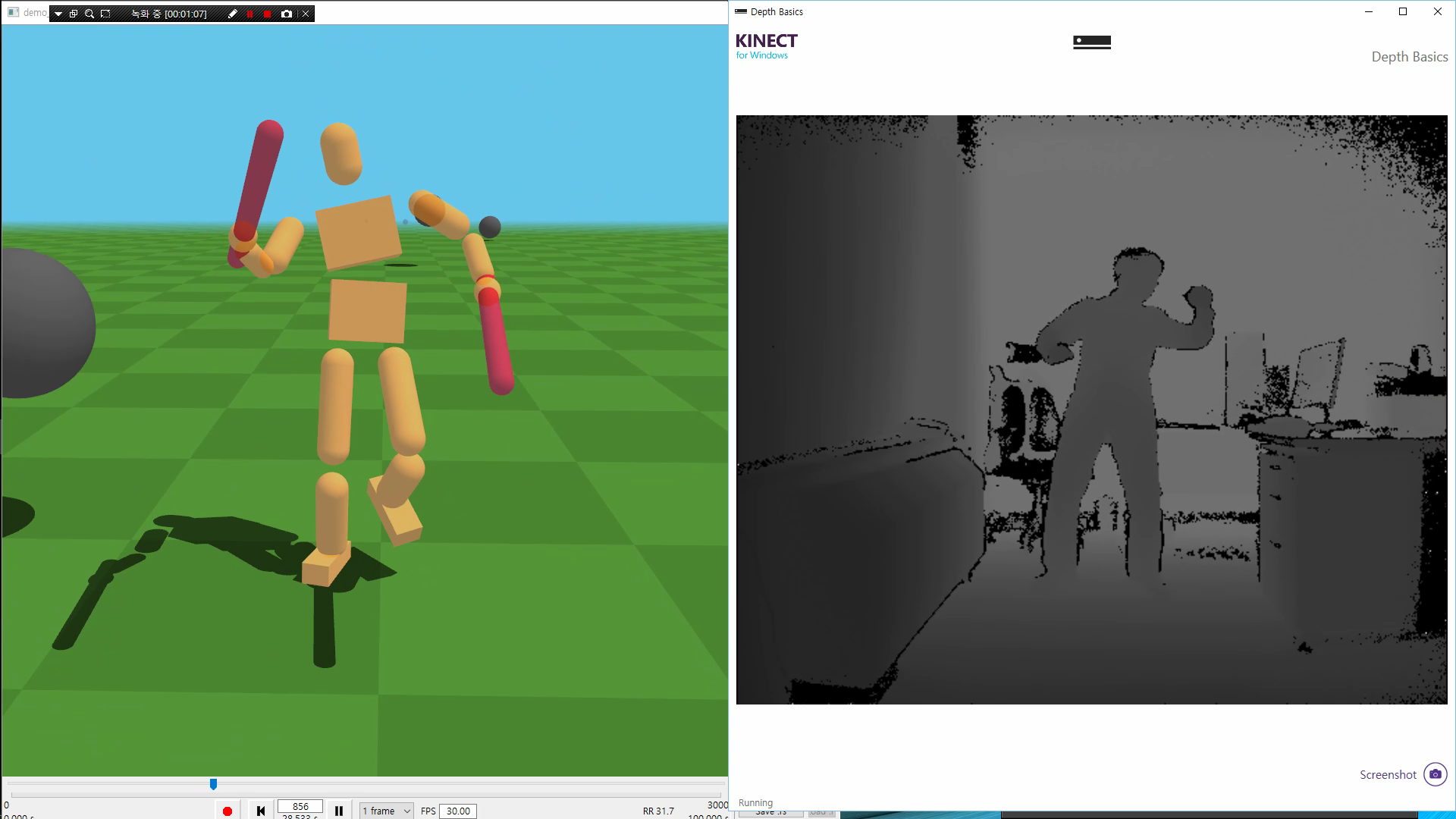}}

		    \includegraphics[trim=100 170 820 50, clip, width=.13\linewidth]{images/teaser/1.png}
		    \includegraphics[trim=100 170 820 50, clip, width=.13\linewidth]{images/teaser/2.png}
		    \includegraphics[trim=100 170 820 50, clip, width=.13\linewidth]{images/teaser/3.png}
		    \includegraphics[trim=100 170 820 50, clip, width=.13\linewidth]{images/teaser/4.png}
		    \includegraphics[trim=100 170 820 50, clip, width=.13\linewidth]{images/teaser/5.png}
		    \includegraphics[trim=100 170 820 50, clip, width=.13\linewidth]{images/teaser/6.png}
		    \includegraphics[trim=100 170 820 50, clip, width=.13\linewidth]{images/teaser/7.png}
	\caption{\label{fig:teaser}
        A user controls a physically simulated character by moving his or her body parts.
        Top: The input depth data \textcolor{rev}{of user poses}.
        Bottom: The simulated character imitating the user's actions.
	}
}

\maketitle

\begin{abstract}
We present a technique for controlling physically simulated characters using user inputs from an off-the-shelf depth camera.
Our controller takes a real-time stream of user poses as input, and simulates a stream of target poses of a biped based on it.
The \textcolor{rev}{simulated} biped mimics the user's actions while moving forward at a modest speed and maintaining balance.
The controller is parameterized over a set of modulated reference motions that aims to cover the range of possible user actions.
For real-time simulation, the best set of control parameters for the current input pose is chosen from the parameterized sets of pre-computed control parameters via a regression method.
By applying the chosen parameters at each moment, the simulated biped can imitate a range of user actions while walking in various interactive scenarios.


\end{abstract}

\section{Introduction}

The development of motion sensing input devices has facilitated the advance of performance-based control techniques.
The Wiimote, an accelerometer-based game controller introduced in 2006, was commercially successful and has had a large impact on how games are played.
Recent devices such as Microsoft Kinect and Leap Motion provide users with more freedom of movement and controllability by tracking their bodies and hands.
These devices have been adopted by many games, which offer new levels of interactivity, and enhance the gameplay experience. %

Another important factor for an immersive experience is physical realism. 
Physics simulations of ragdolls, collisions, and explosions are widely adopted in games to allow natural interactions between the entities, the users, and the virtual world.
Controlling physically simulated bipeds is a promising technique because it has the potential to bring physical realism into the motions of game characters, although such a possibility has yet to be explored thoroughly in current games.  %
Combining these two factors, the performance-based control of a simulated biped would be a powerful tool for creating immersive games and VR applications.

However, the output motion from current off-the-shelf depth cameras is somewhat noisy and inaccurate to be directly used for physically simulated characters.
Output motion is particularly problematic for locomotion control because it often requires reliable contact information for both left and right feet, as well as coordinated movement from the whole-body to be accurately captured.
Also, consumer depth cameras are designed to work in a limited space, such as a small room.
Even walking a few steps at normal speed is not easy in the space.

We propose a \textcolor{rev}{performance-based} biped control system that allows users to make their characters walk forward and mimic their actions.
Our controller takes a real-time stream of user poses from a depth camera and synthesizes a stream of target poses based on it.
By tracking the target poses, the biped can imitate user actions while moving forward at modest speed and maintaining balance.
The controller directly transfers the upper-body portion of the input pose to the target pose to reproduce the user's upper-body movement.
At the same time, our lower-body controller computes the target angles of leg joints which allow the biped to maintain balance and move forward, even though the user is in a small space.
This is a challenging problem because a controller that works for one input motion is not guaranteed to work robustly for a different and unpredictable input motion.
Naturally, users would want to change their pose freely in interactive control scenarios, for example, by bending their knees and changing their upper-body poses to avoid obstacles or to attack enemies.

To deal with this problem, we parameterize our controller over a set of modulated reference motions.
Each modulated reference motion is obtained by editing a single normal walking motion, and a set of modulated reference motions are generated to cover the range of possible user actions.
A set of ten control parameters is optimized for each of modulated reference motions.
During real-time simulation, our system calculates a set of parameters for the current input pose using a regression method with the precomputed controller parameters.
This parameterization enables a stable biped walking under modest but unpredictable variations of user actions.
We demonstrate the effectiveness of our approach by comparing the parameterized controller and the baseline controller without parameterization, and several interactive control demos.

\section{Related Work}

Computer puppetry systems can produce high-quality animations based on a performer's motions in real time. 
A computer puppetry system works by mapping a user pose to a corresponding character pose in an on-line manner using various devices that track human motions.
Some research results focused on producing high quality animations for movies and broadcast performances~\cite{sturman1998,Shin:2001:CPI} using optical or magnetic mocap devices. Such performance-based motion prototyping has been proven effective especially for novice animators who do not have the necessary skills and time for key-framing.
Others studied the potential of low-cost input devices to interactively control human characters~\cite{Oore02adesktop,lee2002,thorne_motion_2004,Jacobson:2014:TMI}. Such approaches were generally better suited for virtual reality applications and game-like scenarios. 
Oore et al. presented a tangible input device and interface for intuitive control of characters having more degrees of freedom than those of the input device~\cite{Oore02adesktop}.
\textcolor{rev}{Yin et al. studied an animation interface that uses a foot pressure sensor pad to interactively control avatars~\cite{yin_footsee:_2003}.}
Terra and Metoyer proposed an approach that allows the user to perform key-frame timing using simple 2D input devices such as a mouse or a stylus~\cite{Terra:2004:PTK}.  Chai and Hodgins employed video cameras and a small set of retro-reflective markers in stereo video, and reconstructed 3D poses using a local linear model to build a low-cost mocap system~\cite{Chai:2005:Performance}. 
Grochow et al. presented an inverse kinematics solver based on a Gaussian process latent variable model (GPLVM) learned from human poses, and demonstrated its robustness using a real-time motion capture system with missing markers~\cite{Grochow:2004:SIK}.
Slyper and Hodgins created a performance animation system that uses five low-cost accelerometers sewn into a shirt. The low-dimensional and noisy signals from the system were supplemented by a database of captured human motions to animate the upper-body of an avatar according to the motion of the performer~\cite{SlyperH08}.  Tautges et al. used four accelerometers attached to the hands and feet of a human actor, and a large number of captured motion sequences to reconstruct the full-body motion of the performer using a lazy-learned neighborhood graph~\cite{Tautges:2011:MRU}.

Using natural user interface to control characters is becoming more widely accepted since the advent of many consumer-grade, real-time motion capture devices such as Microsoft Kinect, Leap Motion, Nintendo Wiimote and many others.  Some researchers focused on creating applications to extend the capabilities of such devices. 
Seol et al. presented a real-time motion puppetry system to animate non-human characters with various skeleton structures using human motion input~\cite{Seol:2013:CFO}.
Rhodin et al. estimated wave properties such as amplitude, frequency, and phase from user motions, and used these properties to robustly control various characters for common cyclic animations from a sparse set of example motions~\cite{Rhodin:2014:IMM,Rhodin:2015}. 
V{\"o}gele et al. presented a method to interactively generate detailed mesh animations using a statistical shape model learned from existing mesh sequences based on inputs from a Kinect sensor~\cite{voegele2012}.

Physics simulation has also been frequently used for interactive character animation, because of its potential for truly responsive and realistic animation~\cite{HodginsWBO95,Laszlo:2000:ICP,Zhao:2005:UII,Yin:2007:SBC,Silva:2008:ISS,Muico:2009,macchietto:2009,Coros:2009:RTC,Lasa:2010:SIGGRAPH,Mordatch:2010:SIGGRAPH,Kwon:2010:CSH,Ye:2010:SIGGRAPH,Wu:2010:SIGGRAPH,Muico:2011:CCP,al_borno_trajectory_2013,ha_iterative_2014,Hamalainen:2015:OCS,2016-TOG-controlGraphs}. The physics-based control of characters is generally regarded as more difficult because of the inherent instability of the floating-base underactuated system. The instability, however, is a key ingredient for reproducing natural responses to perturbations due to changes in the environment.
Many recent publications on physically simulated characters demonstrated dramatic improvements in motion quality, versatility, and robustness. Similar to our work, some of them used feed-back controllers for simulated characters to track captured reference motions~\cite{Sok:2007:SBB,Silva:2008:ISS,Muico:2009,Lee:2010,Kwon:2010:CSH,Lee:2014:LCM,2016-TOG-controlGraphs}. 
However, these controllers were demonstrated using only low DOF input devices such as keyboards and joysticks, and it is not clear whether the results can be generalized to high-dimensional, free-form inputs from natural user interfaces.

Our work is most closely related to approaches for constructing reusable parameterized controllers.
Regression models have been used to map between states, tasks and control parameters, and allowed the development of robust controllers for bipeds walking over terrain~\cite{2012-TOG-TerrainRunner} and performing various stunts~\cite{2016-TOG-controlGraphs} \textcolor{rev}{and quadrupeds jumping over obstacles~\cite{coros_locomotion_2011}} as well as humanoid robots~\cite{daSilvaICRA14,HaICRA2016}.
We also use a similar method for performance-based control.

Below, we briefly survey prior works relevant to performance-based control of simulated avatars.
Ha et al. demonstrated that
a wide range of natural motions can be reconstructed using ground reaction forces from force sensors and arm motions from a hand tracking device~\cite{conf/sca/HaBL11}.
Ishigaki et al. introduced a performance-based control interface for creating physically-plausible motions that preserve the style of example motions using dynamic simulation of a simplified rigid body model~\cite{Ishigaki:2009:PBC}. 
However, the model lacked the ability to maintain balance, and the pitch and roll were controlled using an imaginary force when necessary.
Nguyeny et al. proposed a framework that unifies kinematic playback of motion capture and dynamic simulation for interactively controlling complex interactions~\cite{NguyenyWBPLZ10}.
However, a trade-off between physical correctness and kinematic control must be made. 
\textcolor{rev}{Vondrak et al. recovered biped controllers capable of implicitly simulating the human behavior observed from video and replaying it under other environments~\cite{vondrak_video-based_2012}.}
Shiratori et al. used  
two sets of Wiimotes, and mapped the amplitude, phase and frequency of the sensor inputs to a physically simulated character~\cite{Shiratori:2008:AUI}.
Three performance interfaces (legs, wrists and joystick interface)
were tested on a set of tracks containing turns and pits.
Our goal is different from this work in that a user is provided with direct, free-form control of upper-body motion, as well as indirect control of the lower-body based on the motion of the performer.

\section{Controller}
\label{sec:controller}

\begin{figure}
   \centering
   \includegraphics[trim= 0 0 0 0, clip, width=.8\columnwidth]{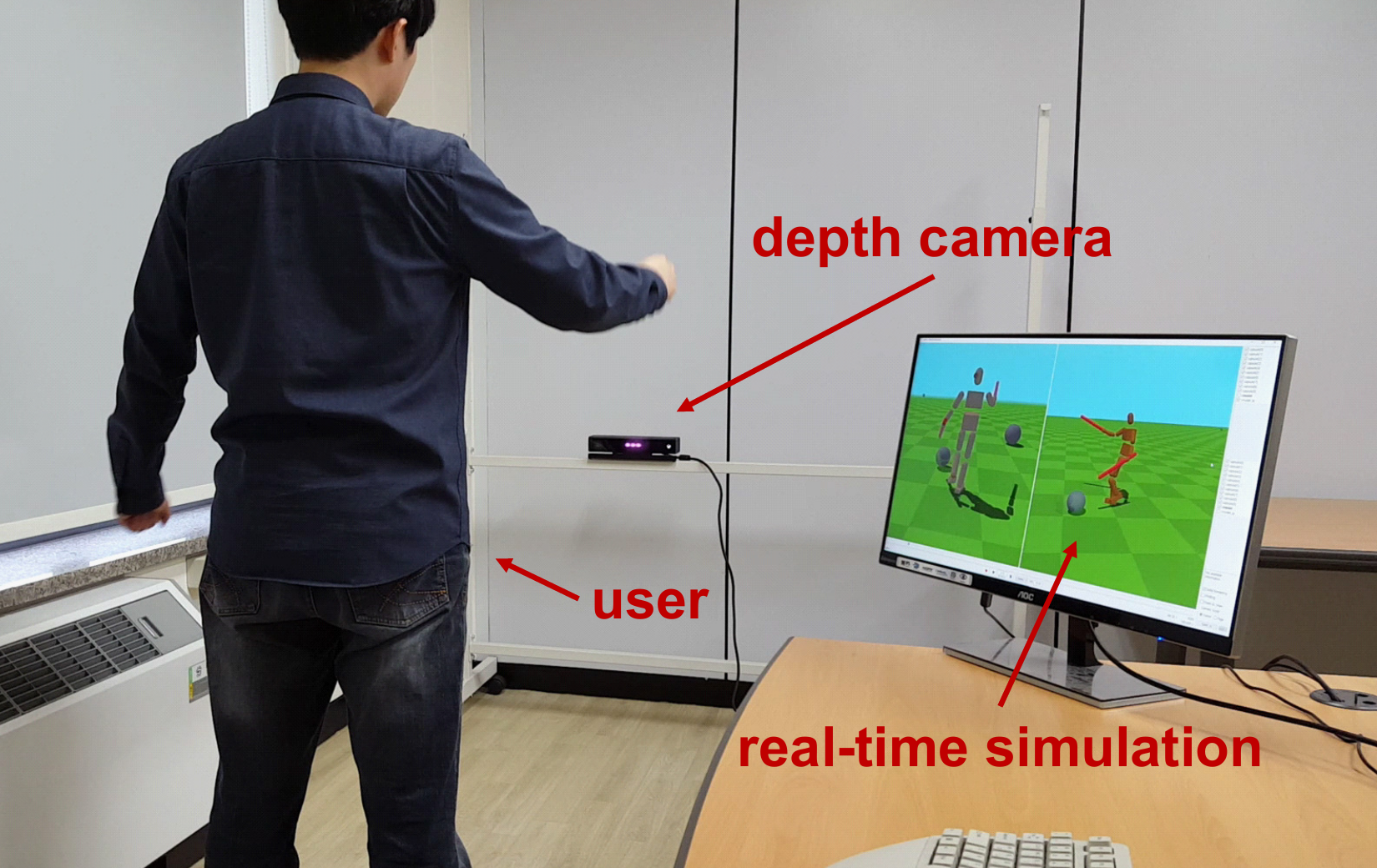}
   \caption{\label{fig:real-time-simul} The depth camera in front of the user captures the user's pose and our system controls a simulated biped as shown on the monitor.}
\end{figure}

Our controller takes a stream of real-time user poses from a depth camera as input (Figure~\ref{fig:real-time-simul}).
Based on the input poses, it synthesizes a stream of target poses and calculates joint torques to track the target poses.
The target poses are generated at the same rate as the input ($30$ Hz), and the joint torques are computed at a higher rate ($900$ Hz). 
A biped character is physically simulated by integrating the joint torques using a forward dynamics simulator.

The target poses need to be generated in a way that the biped can mimic user actions while walking and maintaining balance simply by tracking them.
To build a target pose from each input pose, %
the upper-body joint orientations are first copied from the input pose into the target pose for freeform manipulation of the upper-body.
The time-varying upper-body poses can cause  changes in the center of mass (CM) position of the character and possible instability associated with the changes.
Thus, the remaining degrees of freedom are used for balancing the character. 

Specifically, a stepping strategy is used to regulate the CM of the character. 
We choose to automatically generate the stepping motions based only on the captured pelvis height, and ignore other aspects of the captured lower body motion. This simple approach works well, and the user does not need to walk around the limited capture space from the user-interface point of view.
Some users may prefer to directly use the captured whole-body motion, but the inherent discrepancies in the underlying dynamics between the user and the rigid-body biped model would make the control problem unnecessarily difficult. Also, the contact information from the input data is not reliable due to the input noises.
Even a small delay in foot contact timing can easily disrupt the locomotion stability.
There can be some noises in the upper-body motion captured from the depth camera, but they have relatively small effects on simulation stability in our experiments. 

\begin{figure}
   \centering
   \includegraphics[trim= 0 0 0 0, clip, width=1.\columnwidth]{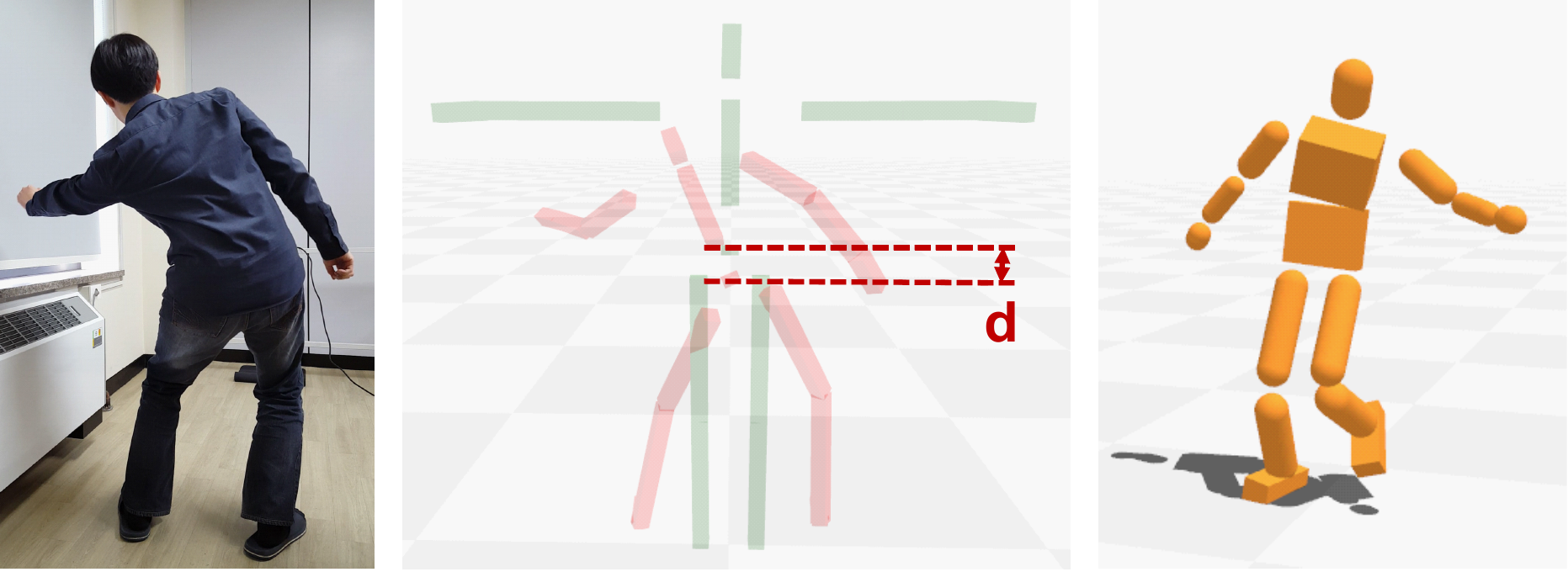}
	\caption{\label{fig:input-pose} The user pose (left) is converted to the target pose (right). 
       Left: A user is bending his knees while moving his arms. 
       Center: The difference in the pelvis height~(d) of the user pose (red) and the upright-standing pose (green) is applied to the lower-body part of the target pose.
	   Right: The resulting target pose makes the simulated biped mimic the user's upper-body pose and knee-bending while walking.   
	   }
\end{figure}

Our lower-body control algorithm closely resembles an existing algorithm proposed by Lee et al.~\shortcite{Lee:2010}. 
The controller in~\shortcite{Lee:2010} takes reference motion data as input, and modifies it to maintain balance by using a few feedback rules. It also synchronizes the contact timings of the reference and the simulated motions.
We slightly modify the controller so that the knee bending angles can be adjusted to imitate a user.
A captured normal walking motion is used as the reference motion for the lower-body part of the character (\textit{base motion}).
To reflect a user's knee bending, we measure the height of the pelvis from the ground for each real-time input pose (Figure~\ref{fig:input-pose}).
The difference between the pelvis height of the input pose and the precomputed upright-standing pose is applied to the current pose of the \textit{base motion} using an analytic inverse kinematics solver.
Note that we use the same skeletal structure for the captured reference pose and the input pose, and thus applying the offset to the reference pose does not cause any problems.
Then, the target pose is built by combining the upper-body portion of the input pose and the lower-body portion of the current lower-body reference pose.
As a result, if a user bends his or her knees, the biped similarly bends its knees while walking.
Once a target pose is generated, our controller calculates all the joint torques for tracking the target pose.
The algorithm for computing the joint torques is almost identical to that in~\shortcite{Lee:2010}. 

\section{Parameterization}

Our controller uses ten feedback parameters $\mathbf \theta$ that determines how to move each leg in a given situation to maintain balance~\cite{Lee:2010}.
Although one set of the parameter values works robustly for small variations in the reference motion to be tracked, the parameters needs to be adjusted depending on the style of the reference motion.
For example, the parameters are hand-tuned for each reference motion in the original work.

However, we cannot use the same approach because there is no fixed reference motion in our case.
The user in front of a depth camera can move freely, for example, swing his or her arms, and the next action is unpredictable.
Even though the control algorithm takes into account the changes in  CM position due to the upper-body movement, 
a single set of control parameters cannot cover enough variations in the upper-body poses.
A set of hand-tuned or optimized parameters for one reference action works well only for user actions that are similar to the reference action.
To deal with a wider range of input poses, we parameterize our controller so that a parameter set corresponds to a style of user action.
We prepare a repertoire of possible user actions, and assign a parameter set for each style.

\subsection{Parameters}

\begin{figure}
   \centering
   \includegraphics[trim= 0 0 0 0, clip, width=1.\columnwidth]{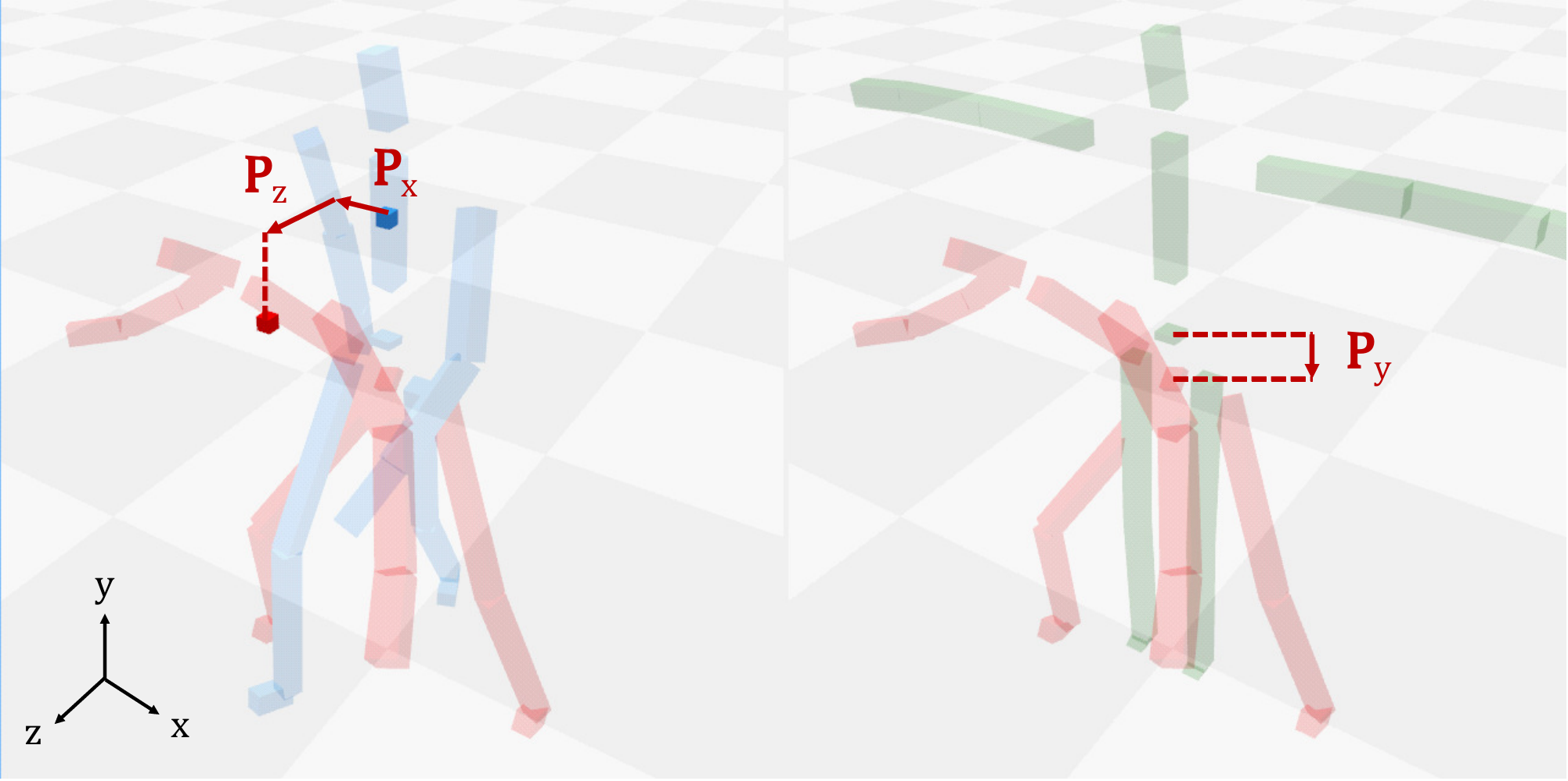}
   \caption{\label{fig:parameters} The parameters for controller parameterization. 
       Left: $\mathbf P_x$ and $\mathbf P_z$ are the horizontal offset of the upper-body CM of the input pose (red cube) from that of the current pose of the base motion (blue cube).
       Right: $\mathbf P_y$ is the offset of the pelvis height of the input pose (red) from that of the standing pose (green).
   }
\end{figure}

Parameters $\mathbf P$ for controller parameterization should be carefully selected to compactly represent the possible space of user actions.
The dimensionality of the parameter space should be low to avoid the curse of the dimensionality problem that occurs when there are insufficient training data.
In our case, the process of preparing the training data is time-consuming because it involves a complex non-linear optimization. 

We choose the following parameters (Figure~\ref{fig:parameters}):
\begin{itemize}
	\item \textbf{Horizontal offset of the upper-body center of mass (CM) ($\mathbf P_x$, $\mathbf P_z$)} 
		of the user with respect to that of the current pose of \textit{base motion} aligned horizontally with the pelvis position of the user.
        It aims to summarize the user's various upper-body actions in terms of balance control.
        The horizontal CM position is important for balanced walking because it directly affects footstep locations.
        We use only the CM of the upper-body parts because the lower-body CM of the simulated biped is mostly determined by the lower-body controller.
		The $xy$-plane is the lateral plane, and the $yz$-plane is the sagittal plane of our simulated biped.
        \\
    \item \textbf{Vertical offset of the pelvis ($\mathbf P_y$)} of the user with respect to the pelvis height of the upright-standing pose.
        It aims to compactly represent the user's lower body motion.
        As stated in the previous section, our lower-body controller can adjust the pelvis height of the biped according to that of the user.
\end{itemize}

A parameter set \textcolor{rev}{$\mathbf P^i=(\mathbf P^i_x, \mathbf P^i_y, \mathbf P^i_z)$} represents a reference motion of which upper-body CM and pelvis height are expressed by \textcolor{rev}{$\mathbf P^i$} (Figure~\ref{fig:parameter-space}).
\textcolor{rev}{To find a set of optimal feedback parameters $\mathbf \theta^{i*}$ for an arbitrary $\mathbf P^i$, each reference motion that corresponds to $\mathbf P^i$ is constructed starting from the \textit{base motion}.}
Specifically, we modify \textcolor{rev}{the joint orientations} of the \textit{base motion} to match a given \textcolor{rev}{$\mathbf P^i$}.
We apply a Jacobian transpose inverse kinematics method \textcolor{rev}{to upper-body joints} to adjust the upper-body CM position of a reference pose, 
and the orientations of leg joints are modified to vertically offset the pelvis via an analytic inverse kinematics solver.
This editing process is conducted for every frame of the \textit{base motion} and the resulting reference motion represents the given \textcolor{rev}{$\mathbf P^i$}.

\subsection{Sampling the Parameter Space}
\label{sec:parameterization_space}

To prepare sets of controller parameters $\mathbf \theta$, we first need to sample the space of parameter $\mathbf P$.
A regular grid of points in the space is created to synthesize a set of reference motions.
An optimization process searches for the best feedback parameters $\mathbf \theta^{i*}$ for each point $\mathbf P^i$, which is described in the next section.

The size of the sampling grid defines the range of possible user actions to be covered by our controller.
We need to choose a reasonable range for each dimension of $\mathbf P$.
Large values of $\mathbf P_x$, $\mathbf P_y$, and $\mathbf P_z$ correspond to walking with excessive knee bending and upper-body stooping, which is unlikely to appear in the user input.

\begin{figure}
   \centering
   \includegraphics[trim= 0 0 0 0, clip, width=1.\columnwidth]{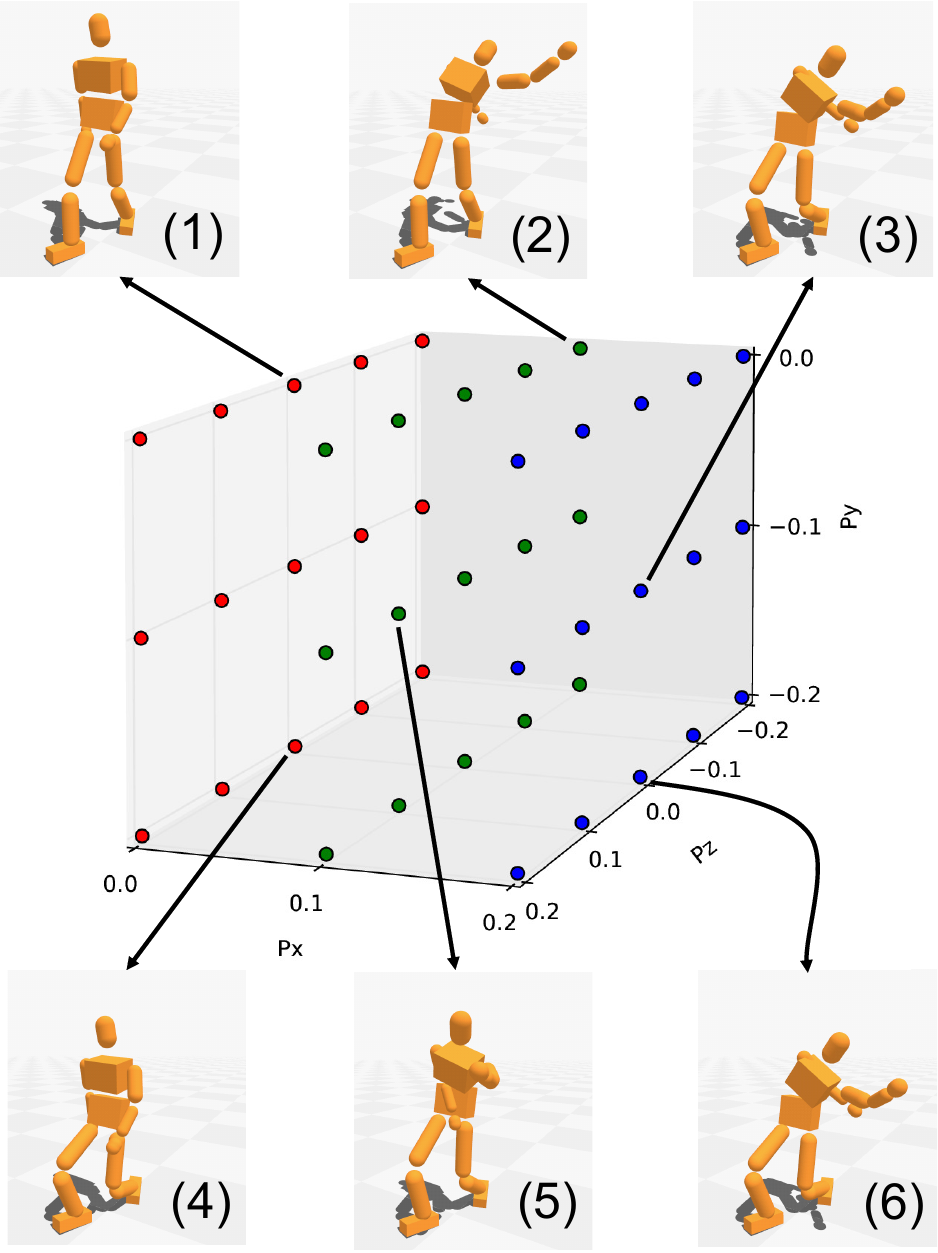}
   \caption{\label{fig:parameter-space} Parameterization space. 
       Each point \textcolor{rev}{$\mathbf P^i$} in the space indicates a reference motion used in the optimization process.
       The modulated reference poses \textcolor{rev}{(1), (2), (3), (4), (5), and (6)} correspond to the first frames of the reference motions indicated by
       the sample parameters 
       \textcolor{rev}{$\mathbf P^1=(0,0,0)$,
       $\mathbf P^2=(0.1,0,-0.2)$,
       $\mathbf P^3=(0.2,-0.1,0)$,
       $\mathbf P^4=(0,-0.2,0)$,
       $\mathbf P^5=(0.1,-0.1,0.1)$,
       and $\mathbf P^6=(0.2,-0.2,0)$}, respectively.
       The points are rendered using three different colors based on their $x$ values for a clean visualization.
   }
\end{figure}

We use $[0, 0.2]$, $[-0.2, 0]$, and $[-0.2, 0.2]$ (in the meter unit) for the ranges of $\mathbf P_x$, $\mathbf P_y$, and $\mathbf P_z$, respectively (Figure~\ref{fig:parameter-space}).
The range of $\mathbf P_x$ contains only non-negative values because we use the same set of feedback parameters for a negative-valued $\mathbf P_x$ which indicates that the lateral leaning of the upper-body has the same magnitude but in the opposite direction.
This greatly reduces the number of samples, and shortens the computation time.
We sample three points in the x and y directions, and five points along the z direction, and thus the total number of points is $3 * 3 * 5 = 45$.

We first sort the $45$ sample points based on the Euclidian distance from the origin $\mathbf P^0 = (0, 0, 0)$.
Starting from the nearest point to $\mathbf P^0$, we sequentially optimize the sets of feedback parameters $\mathbf \theta^i$ for each grid point $\mathbf P^i$ in the sorted order.
Finding robust controller parameters for $\mathbf P^i$ farther from $\mathbf P^0$ is more challenging than for those that are closer to $\mathbf P^0$ because the former requires walking with a more pronounced upper-body lean and crouched knees.
To improve the optimization performance, we provide initial guesses $\mathbf \theta^{i0}$ for each optimization of $\mathbf P^i$ using other parameters $\mathbf \theta^{j*}$ that have already been optimized.
We choose an initial guess $\mathbf \theta^{i0}$ from $\mathbf \theta^{j*}$ of nearby $k_1$ points via a k-nearest neighbor regression.

Even when the size of the parameterization space is reasonable, the optimizer may fail to find robust feedback parameters for a sample point.
In such a case, we use a heuristic search algorithm that repeatedly scales the distance between the sample point and the origin by $0.8$ until a set of feedback parameters is successfully found.

\section{Optimization}

Given a new reference motion represented by $\mathbf P^i$, the controller feedback parameters $\mathbf \theta^i$ are optimized by the objective function proposed in~\shortcite{lee_push-recovery_2015}:
\begin{equation}
    E(\mathbf \theta) = w_1 E_\mathrm{balance} + w_2 E_\mathrm{track} + w_3 E_\mathrm{param},
	\label{eq:objective_func}
\end{equation}
 where $E_\mathrm{balance}$ penalizes the falling down of the biped, $E_\mathrm{track}$ penalizes the deviation of the simulated poses from the reference poses, and $E_\mathrm{param}=\|\mathbf \theta\|$ penalizes large parameter values to prevent abrupt changes in reference motions, which often make the simulation unstable.
We set the weight values $w_1=2000$, $w_2=1.0$, and $w_3=1.0$.

An evaluation of the objective function requires a simulation of 20 seconds in duration:
ten seconds for the given reference motion and ten seconds for the mirrored motion in the sagittal plane of the biped.
Evaluating both the reference motion and the mirrored motion allows us to handle the left and right leaning of the upper-body symmetrically using a single set of optimized parameters $\mathbf \theta^{i*}$.
It also enhances the robustness of the controller by optimizing for the two different initial conditions, starting with the left foot and the right foot.
The result of the objective function is set as the sum of the results for the first ten seconds and the next ten seconds.

We use a Covariance Matrix Adaptation (CMA) method to find a set of optimal parameters $\mathbf \theta^{i*}$.
The size of the CMA population is 24.
The optimization for each $\mathbf P^i$ is terminated when the number of iterations exceeds 200 or the result of the objective function is smaller than 2000, to shorten the computation time.
An evaluated value larger than 3000 is regarded as a failure case, and the heuristic search algorithm is executed as described in Section~\ref{sec:parameterization_space}.

\section{Real-time Simulation}

During the real-time simulation, a user stands up in front of a depth camera and freely move his or her body parts (Figure~\ref{fig:real-time-simul}).
Our controller generates target poses based on the input poses to imitate the user's actions in the simulated environment, as described in Section~\ref{sec:controller}.
Only a carefully selected set of feedback parameters leads to robust walking of a simulated biped.

We compute feedback parameters $\mathbf \theta^{\mathrm{cur}}$ at the input pose frequency.
First, the parameterization parameter $\mathbf P^{\mathrm{cur}}$ is extracted from the current input pose by measuring the position of the upper-body CM and the height of the pelvis.
Then, we compute $\mathbf \theta^{\mathrm{cur}}$ using a multivariate regression of the precomputed controller parameters.
\textcolor{rev}{Specifically, precomputed parameters $\mathbf \theta^{j*}$ of $k_2$ nearest points are interpolated using a k-nearest neighbor scheme with inverse distance weighting (IDW).}
The resulting $\mathbf \theta^{\mathrm{cur}}$ makes the biped imitate user poses while walking and maintaining balance.

\section{Results}

We use a Microsoft Kinect (version 2) in our experiments.
A user input pose is built from global joint orientation data acquired using Kinect for Windows SDK 2.0.
The frame rate of both the input poses for real-time simulation and the reference motions used in the optimization are 30 Hz.
Although a new pose is mostly obtained from the Kinect camera within $1/30$ seconds, we reuse the same input pose from the previous frame when the input pose is delayed occasionally. 

We use GEAR~\cite{gear_library} to simulate our biped character with a simulation time step of $1/900$ seconds.
The number of neighbors for both the initial guess of the optimization ($k_1$) and the real-time interpolated parameters ($k_2$) are $4$.
Using 12 cores on a Xeon-based workstation (Intel X5650@2.67 GHz), the parameter optimization for all 45 sample points takes approximately $35$ hours, including the time spent in the heuristic search algorithm for failure cases.

\begin{figure}
   \centering
   \includegraphics[trim= 0 0 0 0, clip, width=1.\columnwidth]{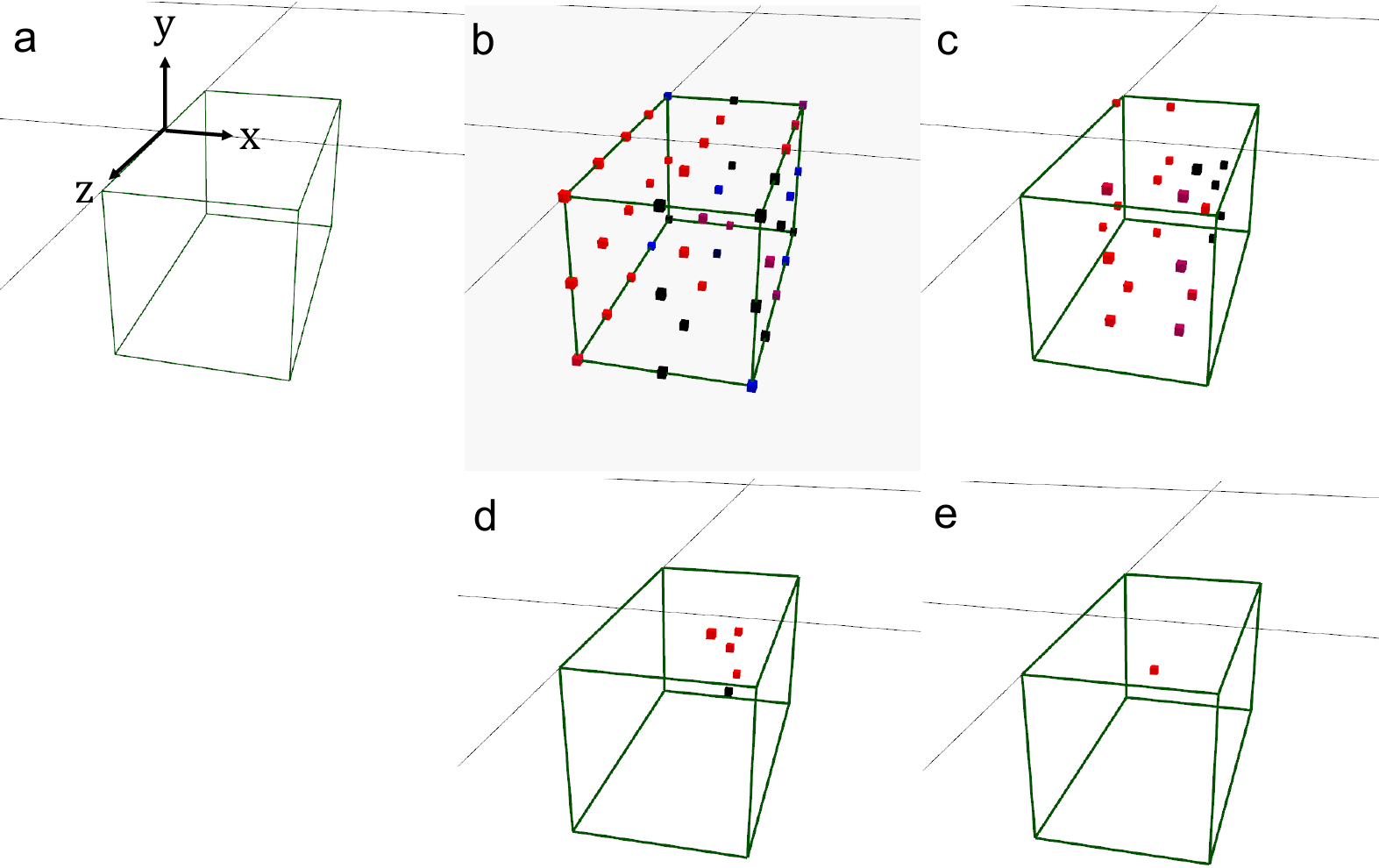}
   \caption{\label{fig:opt-space} Optimization of parameter space. 
       The range of the initial parameter space is rendered as the green cube (a).
       Minimum evaluated values of the first 45 sample points (b), of 21 sample points scaled by $0.8$ (c), of 5 sample points scaled by $0.8^2$ (d), of 1 sample point scaled by $0.8^4$ (e).
        Successful points (evaluated value from $2000$ to $3000$) are rendered in color from red (value of $2000$ or less) to purple (up to value of $3000$) and failed points (evaluated value more than $3000$) are rendered in color from blue (value of $3000$) to black (value of $10000$ or more).
   }
\end{figure}

\begin{figure}
   \centering
   \includegraphics[trim= 0 0 0 0, clip, width=1.\columnwidth]{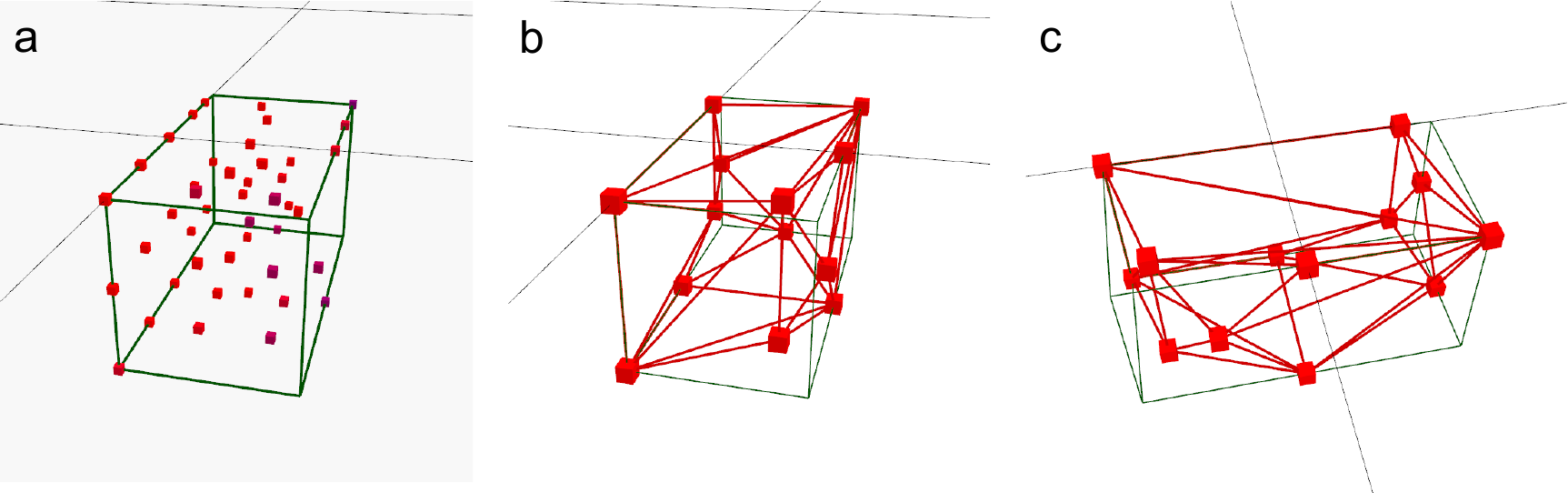}
   \caption{\label{fig:shrink-space} "Shrunk" parameter space. 
       (a) All successful 45 sample points.
       (b)(c) The convex hull of the successful points.
   }
\end{figure}

\textbf{Shape of the Optimized Parameter Space.}
The optimizer tries to find the controller parameters for all sample points in the range of $[0, 0.2]$, $[-0.2, 0]$, and $[-0.2, 0.2]$ of the parameter space,
but sometimes it fails and resorts to the heuristic search algorithm.
Because we repeatedly scaled down the distance between such samples and the origin as stated in Section~\ref{sec:parameterization_space} (Figure~\ref{fig:opt-space}),  
the final parameter space is smaller than the initial one (Figure~\ref{fig:shrink-space}).
We visualize the successful points as a convex hull to show \textcolor{rev}{the overall shape of} the final space that can be covered by our parameterized walking controller.
\textcolor{rev}{Note that the convex hull is used only to visualize the shape of the space, and it does not guarantee a successful simulation of a sample point in it.}
Figure~\ref{fig:shrink-space} shows that our controller is more unstable when the user leans the upper-body backward, or when the user leans and bends knees at the same time.

\textbf{Comparison of Parameterized and Baseline Controller.}
We compared the robustness of our parameterized controller and the controller optimized only for the \textit{base motion} (\textit{baseline controller}).

\begin{figure}
	\centering
        {\transparent{1}
		    \reflectbox{\includegraphics[trim=100 100 150 100, clip, width=.13\linewidth]{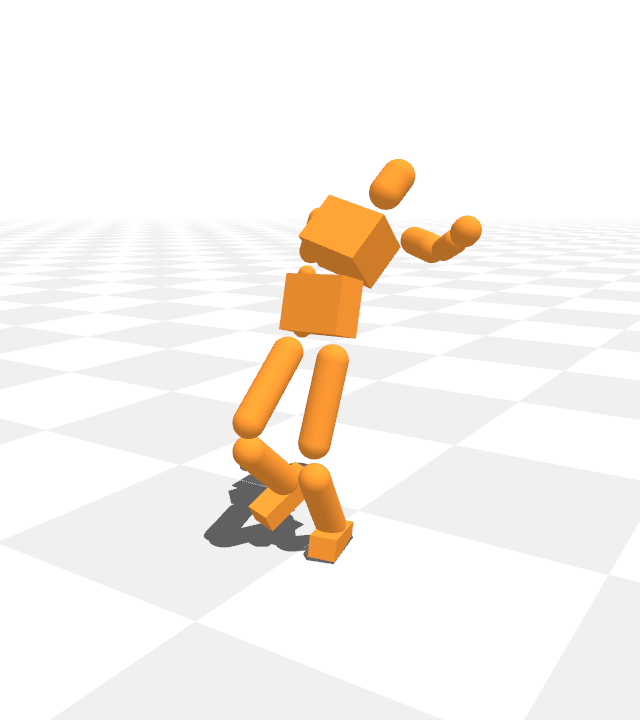}}
		    \reflectbox{\includegraphics[trim=100 100 150 100, clip, width=.13\linewidth]{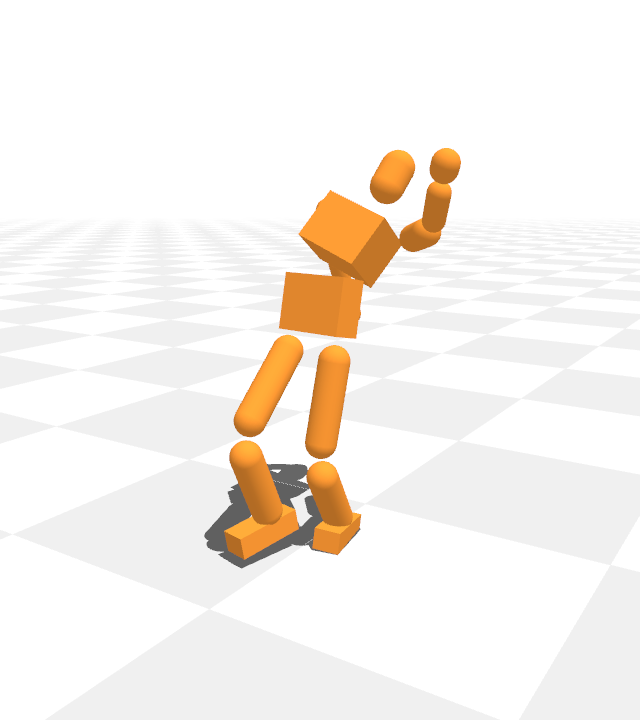}}
		    \reflectbox{\includegraphics[trim=100 100 150 100, clip, width=.13\linewidth]{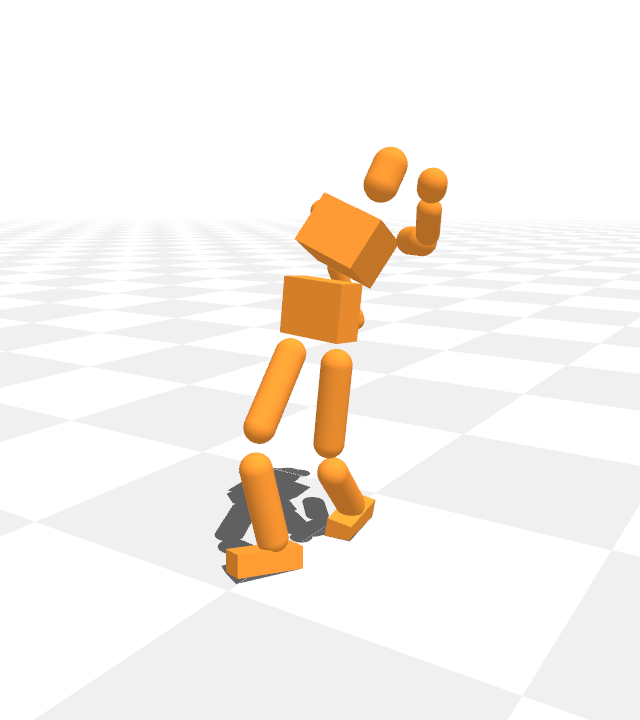}}
		    \reflectbox{\includegraphics[trim=100 100 150 100, clip, width=.13\linewidth]{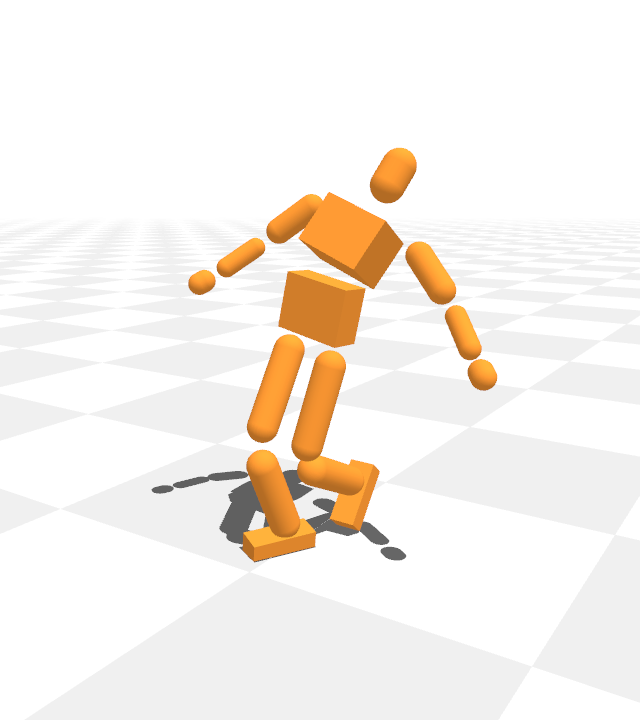}}
		    \reflectbox{\includegraphics[trim=100 100 150 100, clip, width=.13\linewidth]{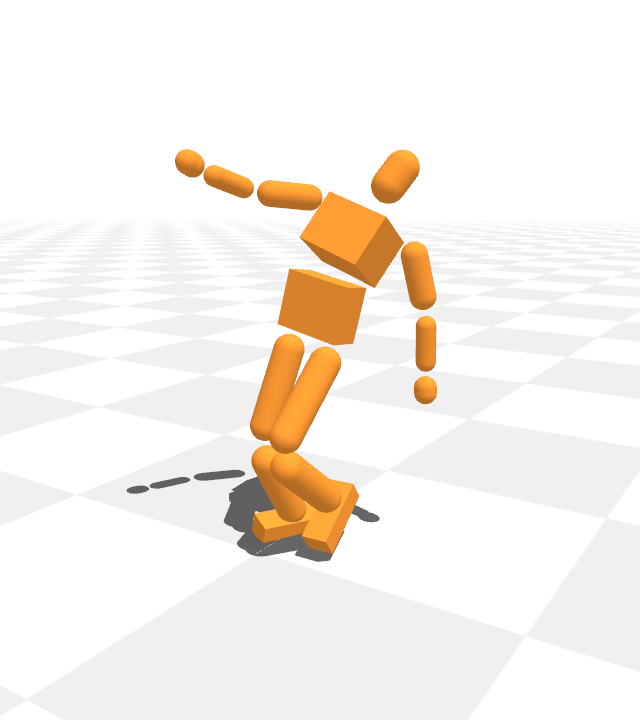}}
		    \reflectbox{\includegraphics[trim=100 100 150 100, clip, width=.13\linewidth]{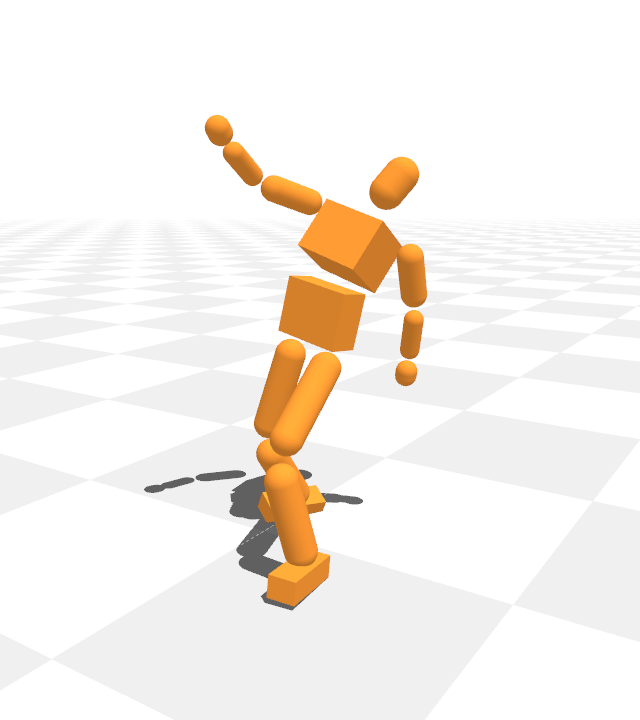}}
		    \reflectbox{\includegraphics[trim=100 100 150 100, clip, width=.13\linewidth]{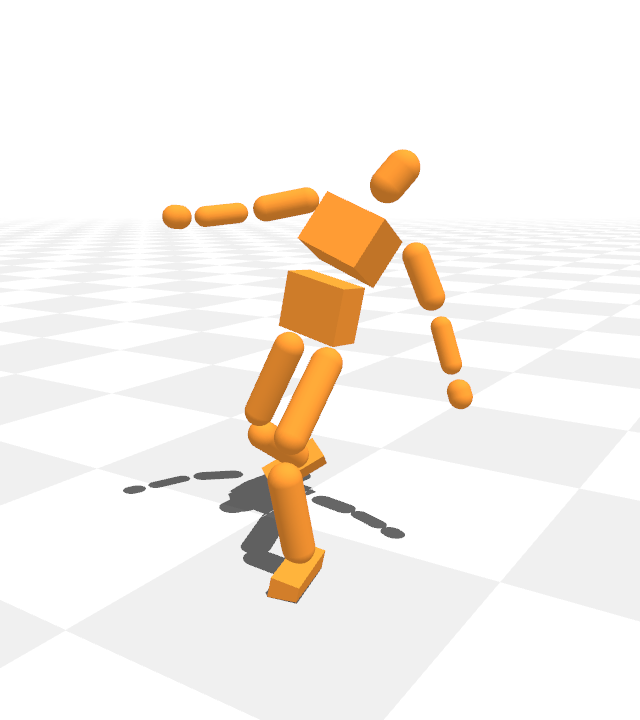}}
        }
        {\transparent{.7}
		    \reflectbox{\includegraphics[trim=100 100 150 100, clip, width=.13\linewidth]{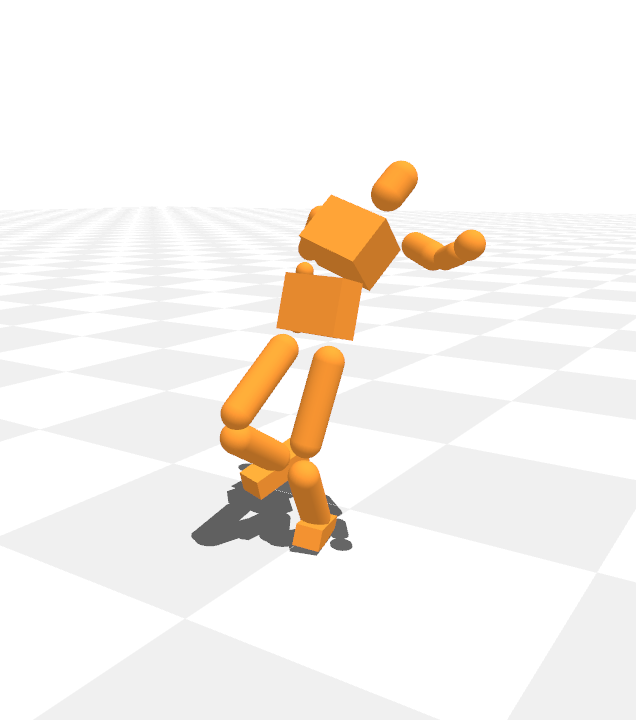}}
		    \reflectbox{\includegraphics[trim=100 100 150 100, clip, width=.13\linewidth]{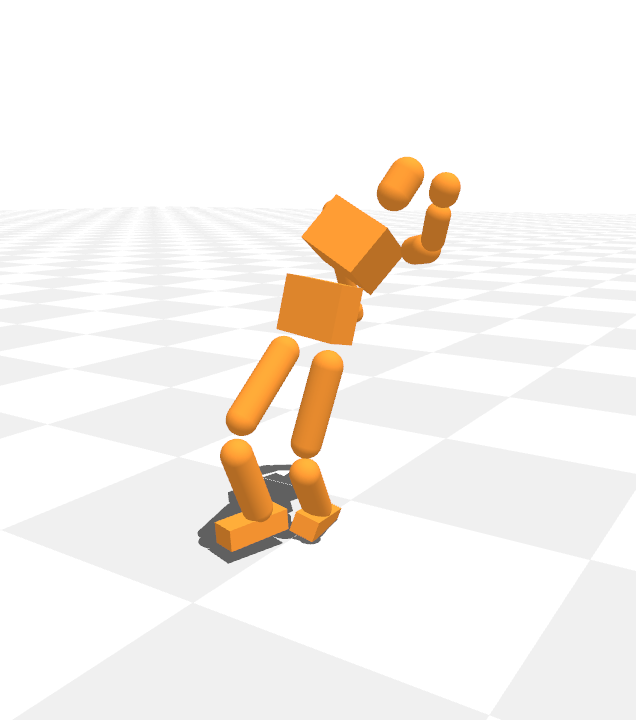}}
		    \reflectbox{\includegraphics[trim=100 100 150 100, clip, width=.13\linewidth]{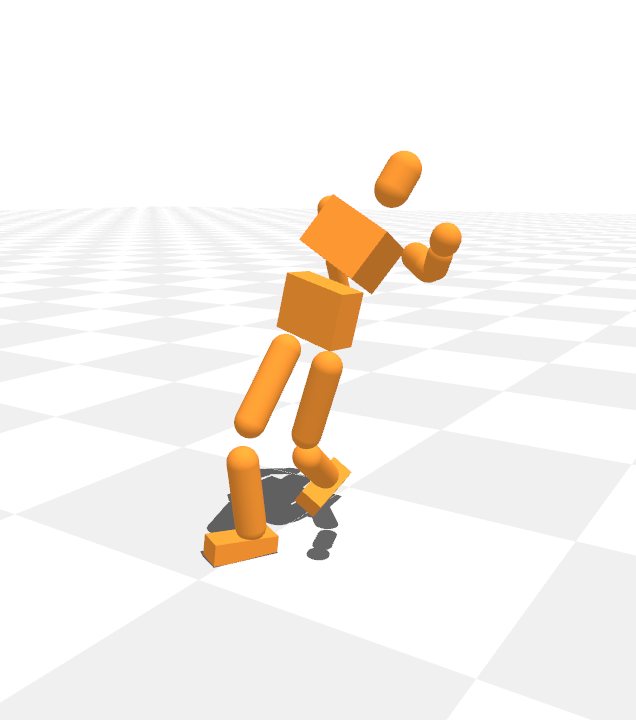}}
		    \reflectbox{\includegraphics[trim=100 100 150 100, clip, width=.13\linewidth]{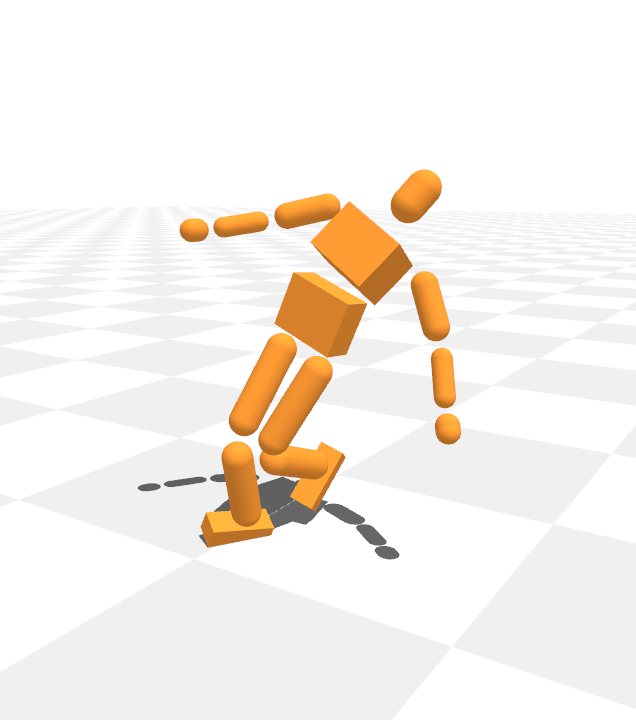}}
		    \reflectbox{\includegraphics[trim=100 100 150 100, clip, width=.13\linewidth]{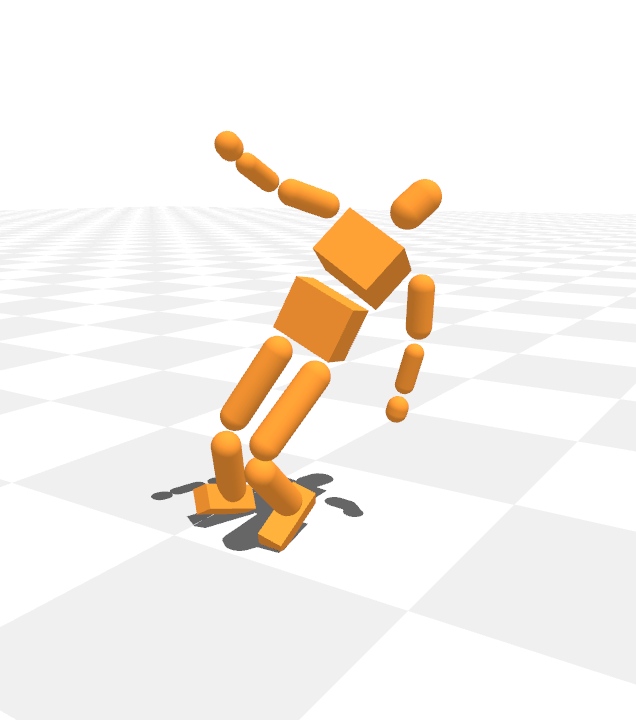}}
		    \reflectbox{\includegraphics[trim=100 100 150 100, clip, width=.13\linewidth]{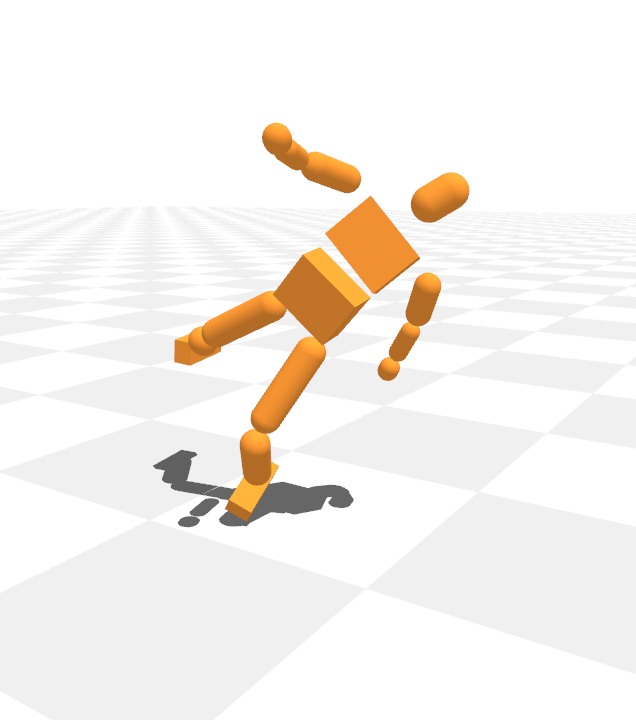}}
		    \reflectbox{\includegraphics[trim=100 100 150 100, clip, width=.13\linewidth]{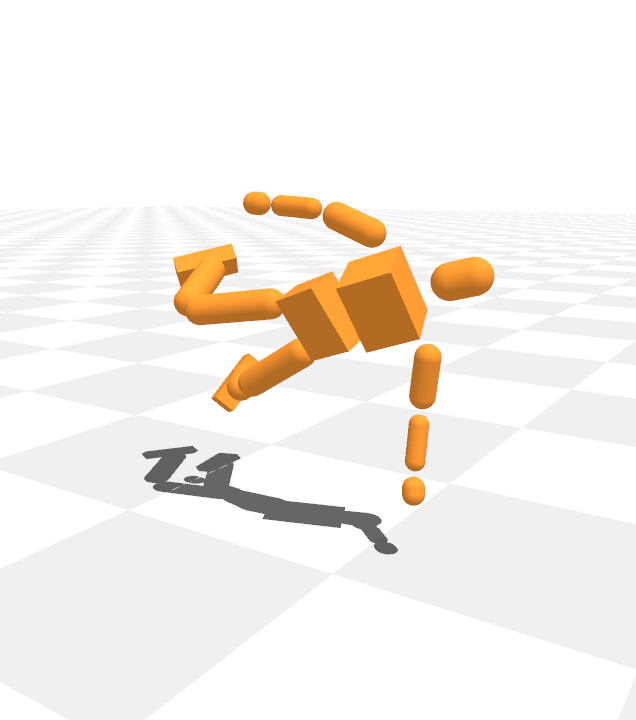}}
        }
        {\transparent{1}
		    \reflectbox{\includegraphics[trim=100 100 150 100, clip, width=.13\linewidth]{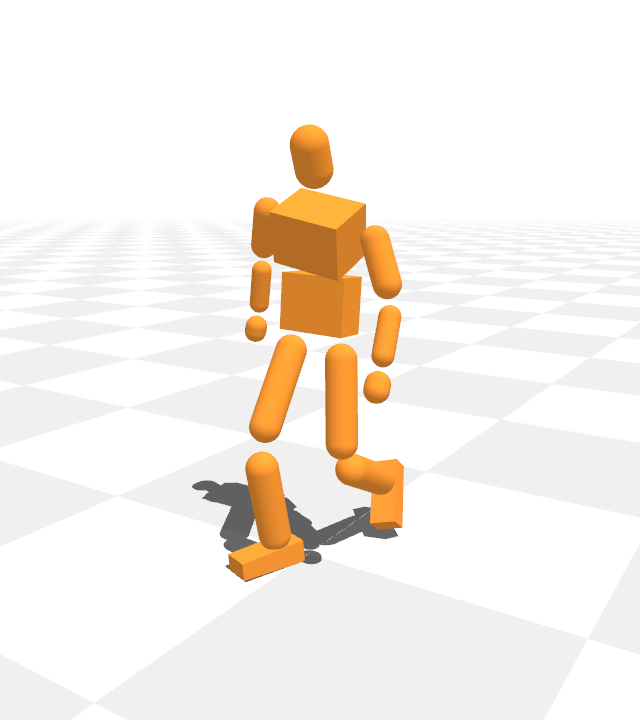}}
		    \reflectbox{\includegraphics[trim=100 100 150 100, clip, width=.13\linewidth]{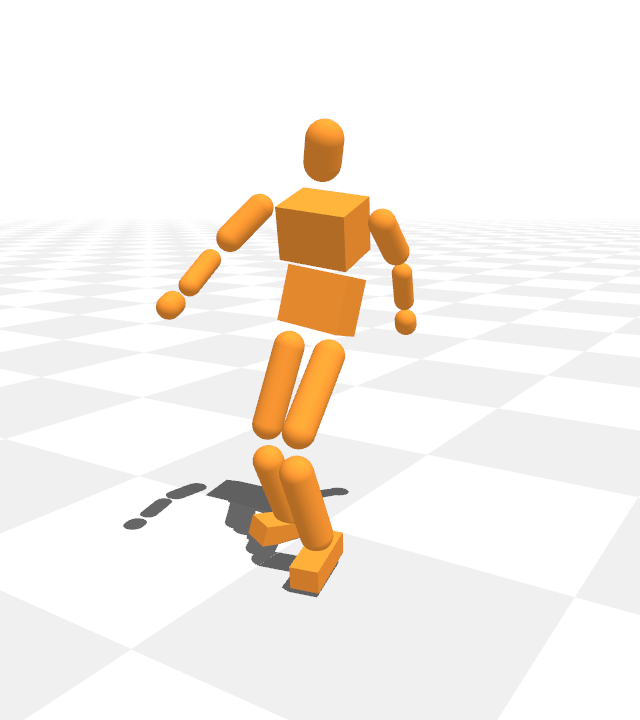}}
		    \reflectbox{\includegraphics[trim=100 100 150 100, clip, width=.13\linewidth]{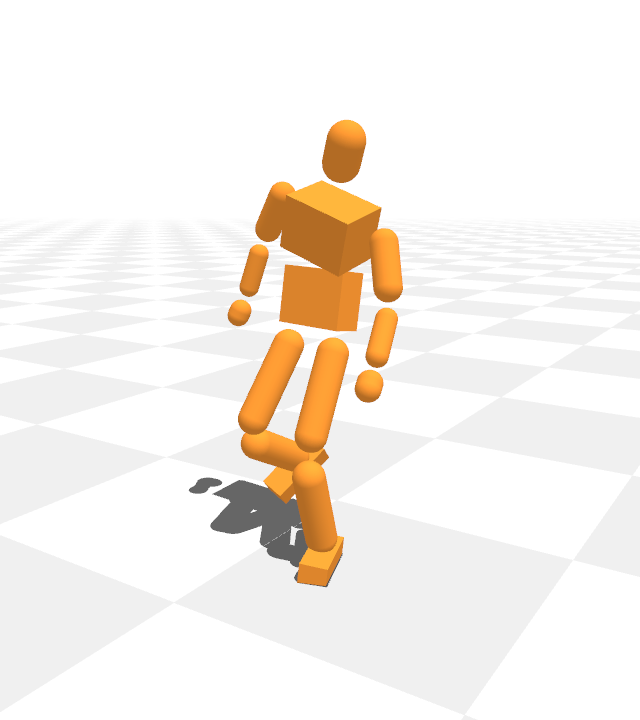}}
		    \reflectbox{\includegraphics[trim=100 100 150 100, clip, width=.13\linewidth]{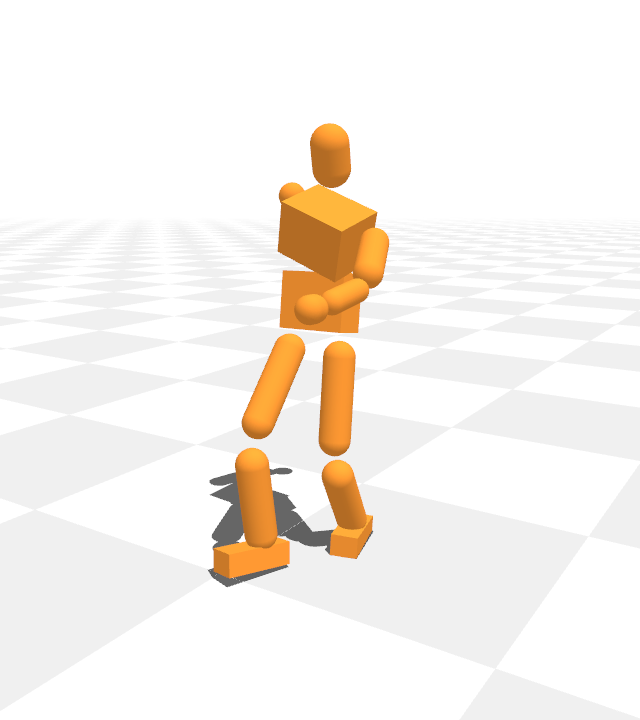}}
		    \reflectbox{\includegraphics[trim=100 100 150 100, clip, width=.13\linewidth]{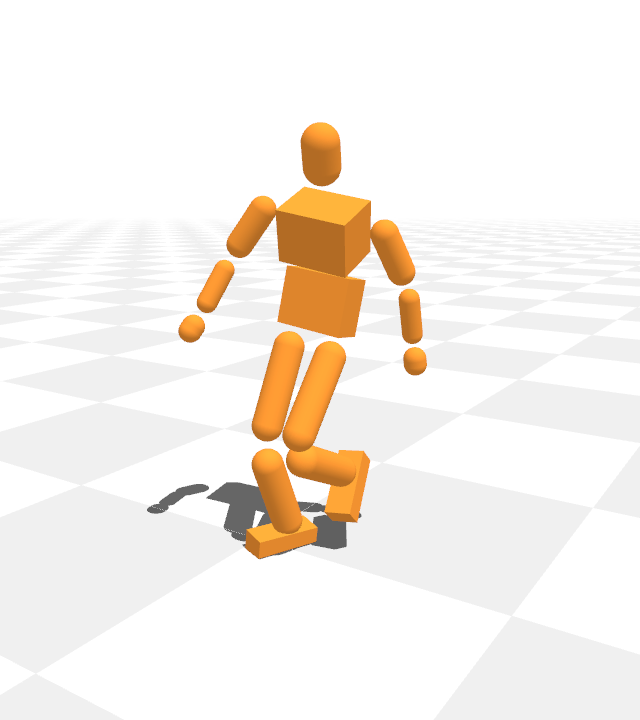}}
		    \reflectbox{\includegraphics[trim=100 100 150 100, clip, width=.13\linewidth]{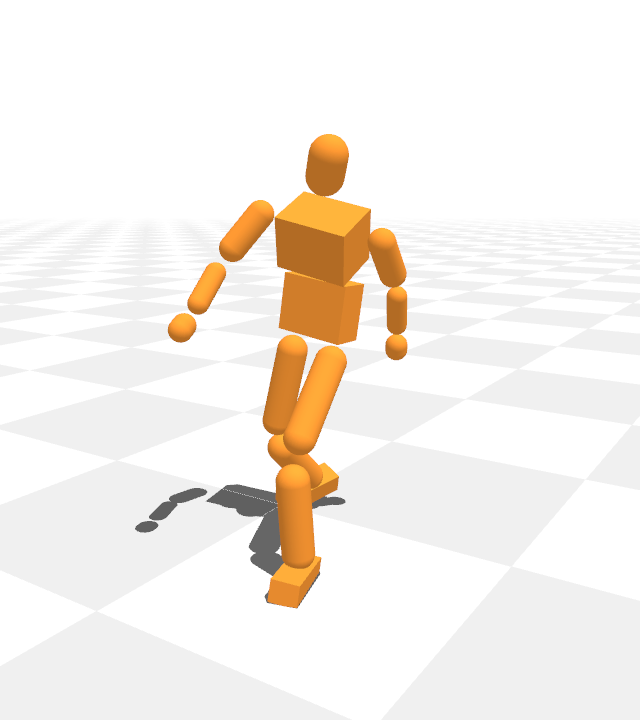}}
		    \reflectbox{\includegraphics[trim=100 100 150 100, clip, width=.13\linewidth]{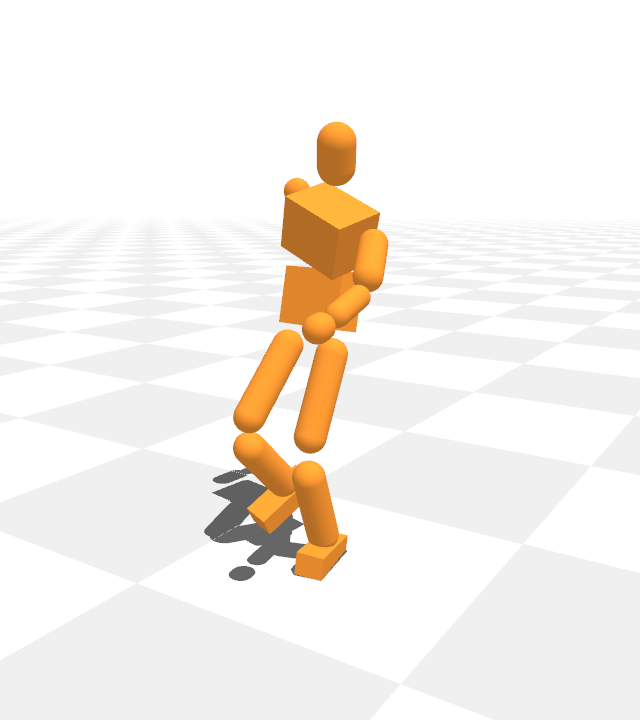}}
        }
        {\transparent{.7}
		    \reflectbox{\includegraphics[trim=100 100 150 100, clip, width=.13\linewidth]{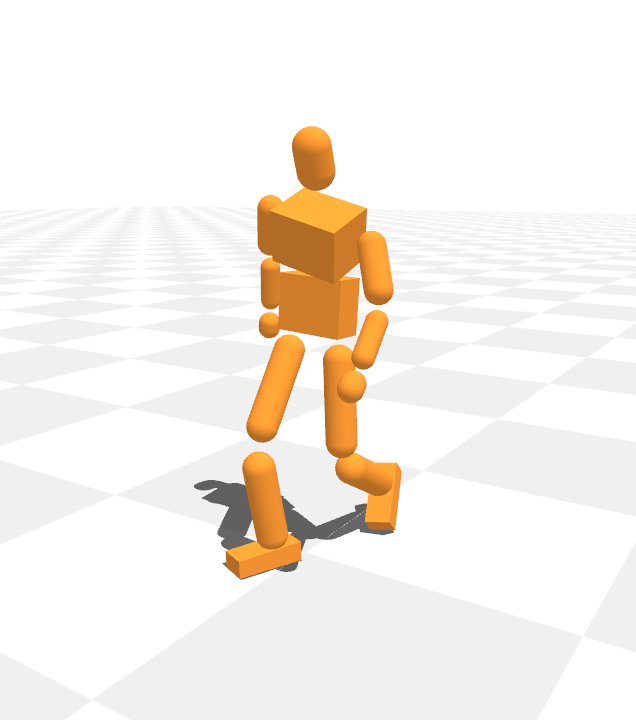}}
		    \reflectbox{\includegraphics[trim=100 100 150 100, clip, width=.13\linewidth]{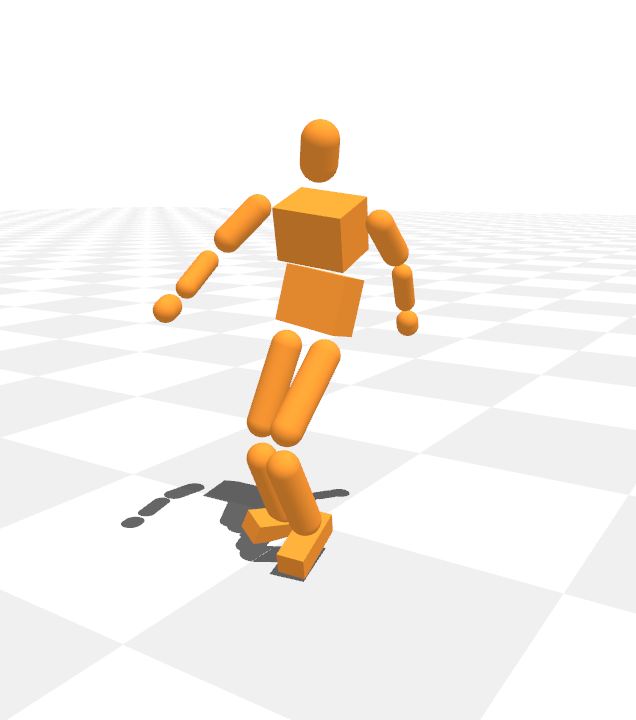}}
		    \reflectbox{\includegraphics[trim=100 100 150 100, clip, width=.13\linewidth]{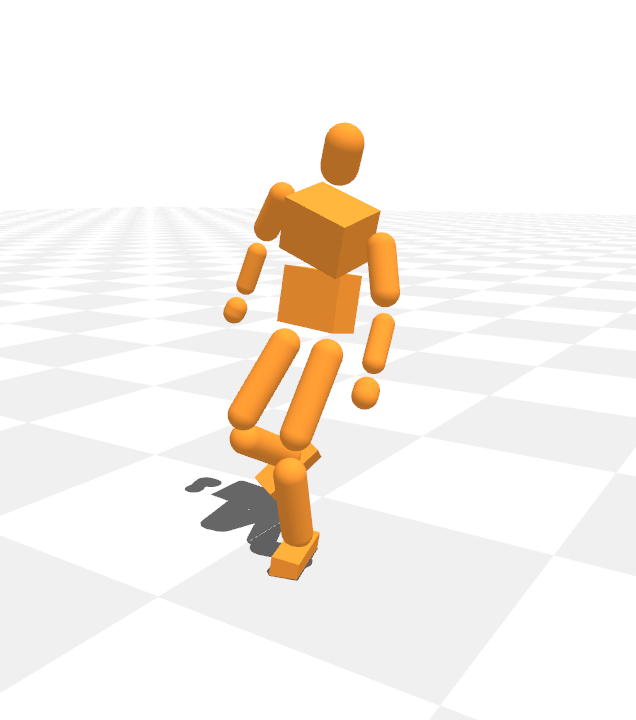}}
		    \reflectbox{\includegraphics[trim=100 100 150 100, clip, width=.13\linewidth]{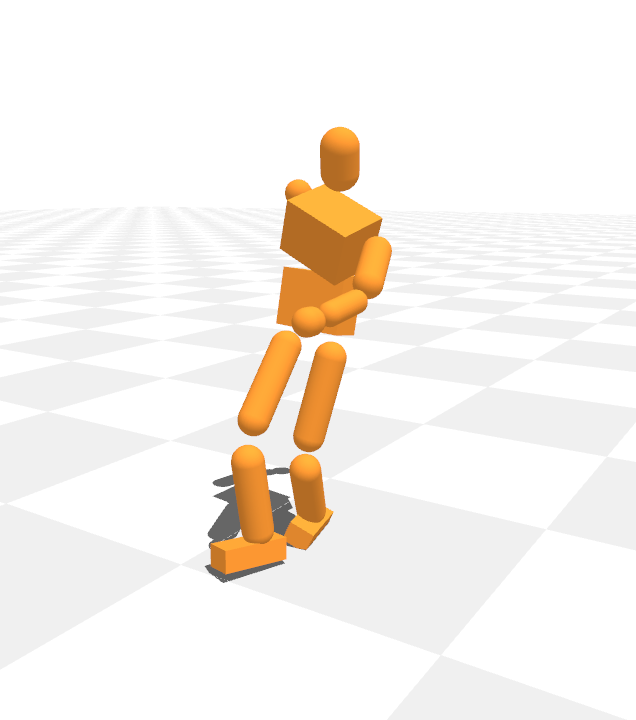}}
		    \reflectbox{\includegraphics[trim=100 100 150 100, clip, width=.13\linewidth]{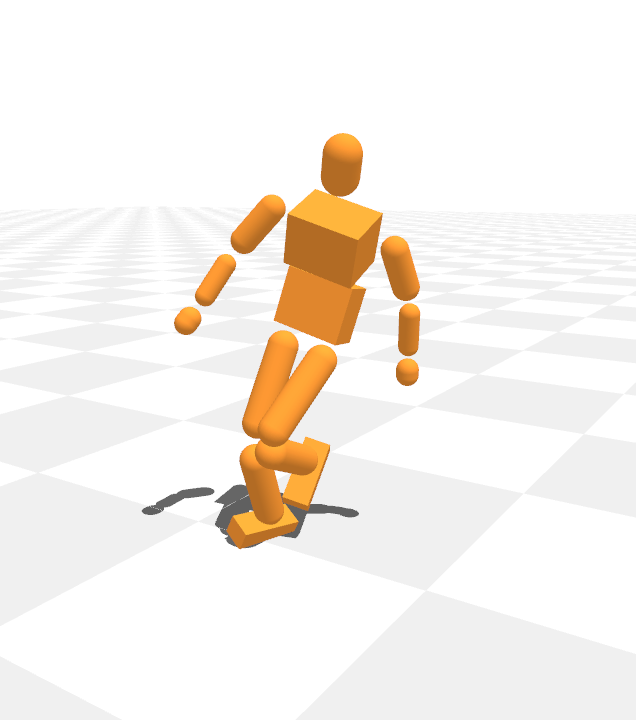}}
		    \reflectbox{\includegraphics[trim=100 100 150 100, clip, width=.13\linewidth]{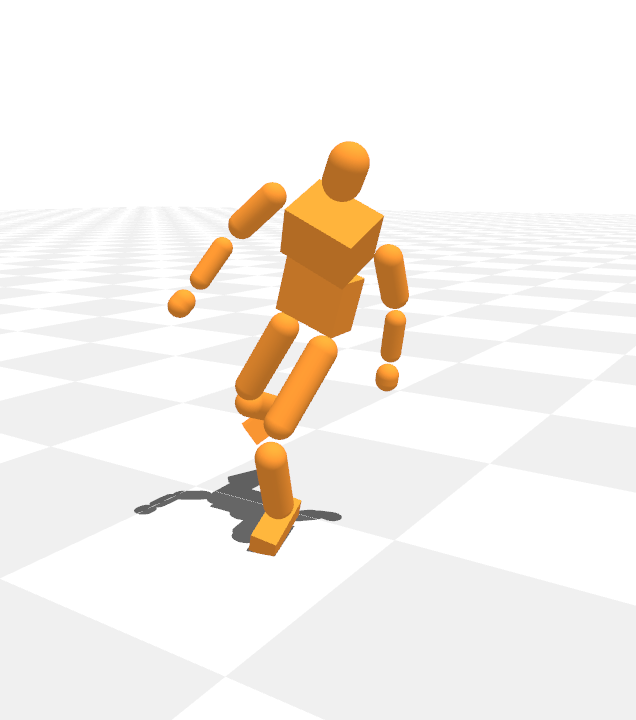}}
		    \reflectbox{\includegraphics[trim=100 100 150 100, clip, width=.13\linewidth]{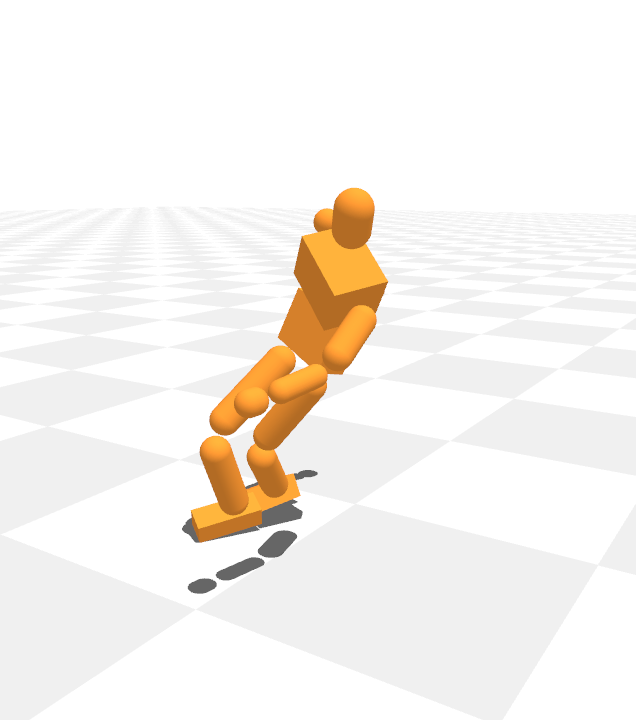}}
        }
	\caption{\label{fig:comp-ik}
            Comparison of the parameterized controller (row 1, 3) and the baseline controller (row 2, 4).
            Row 1, 2: Simulation of the reference motion for $\mathbf P=(0.05, -0.05, -0.15)$.
            Row 3, 4: Simulation of the reference motion for $\mathbf P=(0.05, -0.05, 0.05)$.
	}
\end{figure}

In the first comparison, we used the automatically generated reference motions corresponding to multiple sets of parameters $\mathbf P$ as the input (Figure~\ref{fig:comp-ik}).
The reference motions contained various combinations of upper-body leaning and knee bending.
We found the parameterized controller can deal with a much wider range of modulated reference motions.
For example, the parameterized controller generated stable walking simulation for a motion with $\mathbf P=(0.05, -0.05, -0.15)$, but the baseline controller failed even for a motion with a smaller change: $\mathbf P=(0.05, -0.05, 0.05)$.

\begin{figure}
   \centering
   \includegraphics[trim= 0 0 0 0, clip, width=1.\columnwidth]{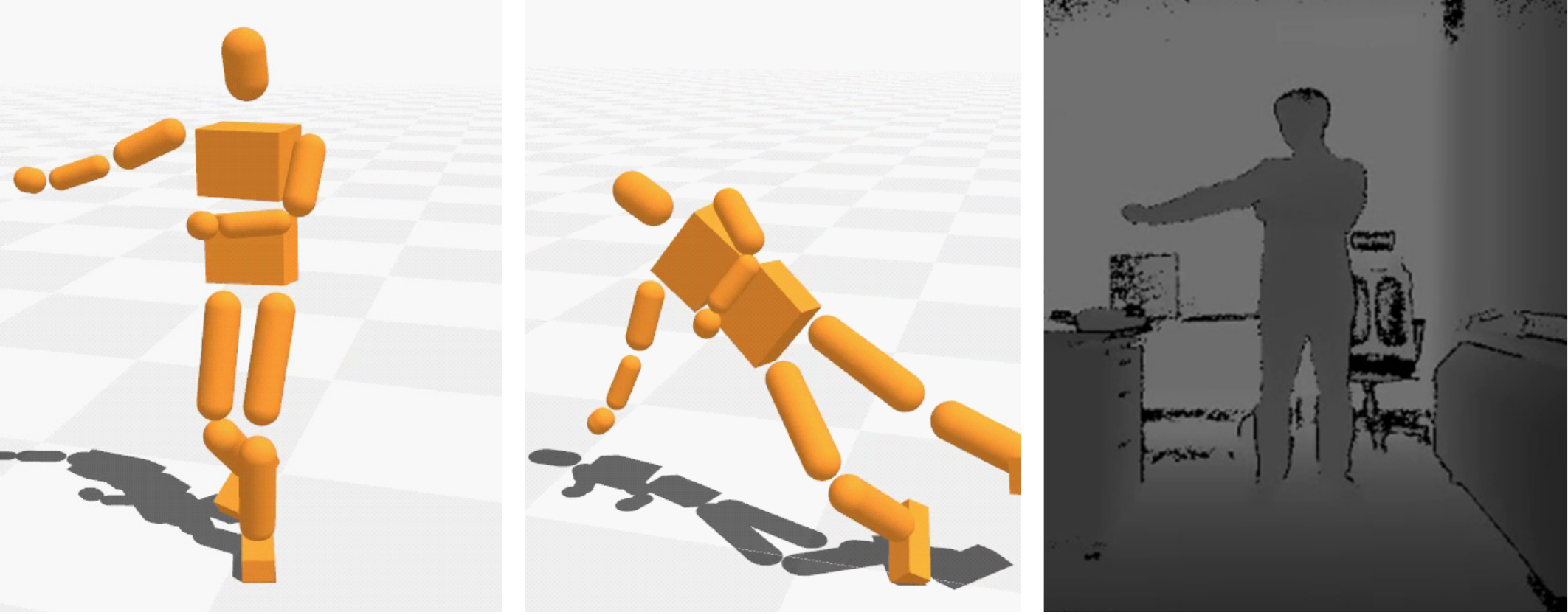}
   \caption{\label{fig:comp-kinect} Comparison of controllers with real-time pose input. 
       Left: The parameterized controller.
       Middle: The baseline controller.
       Right: The input depth data of the real-time user pose.
   }
\end{figure}

Our second comparison used real-time user poses from a depth camera as input (Figure~\ref{fig:comp-kinect}).
The baseline controller easily failed when the user simply moved their arms to the right side or slightly bent their knees.
The parameterized controller maintained balance during the more complex movement of the user's upper-body and knees.

\textbf{Demo Games.}
The performance-based interface and physical simulation can improve immersion in the virtual environment and the gameplay experience.
We designed two simple game demonstrations to make the best use of our performance-based biped controller.

\begin{figure}
   \centering
   \includegraphics[trim= 0 0 0 0, clip, width=1.\columnwidth]{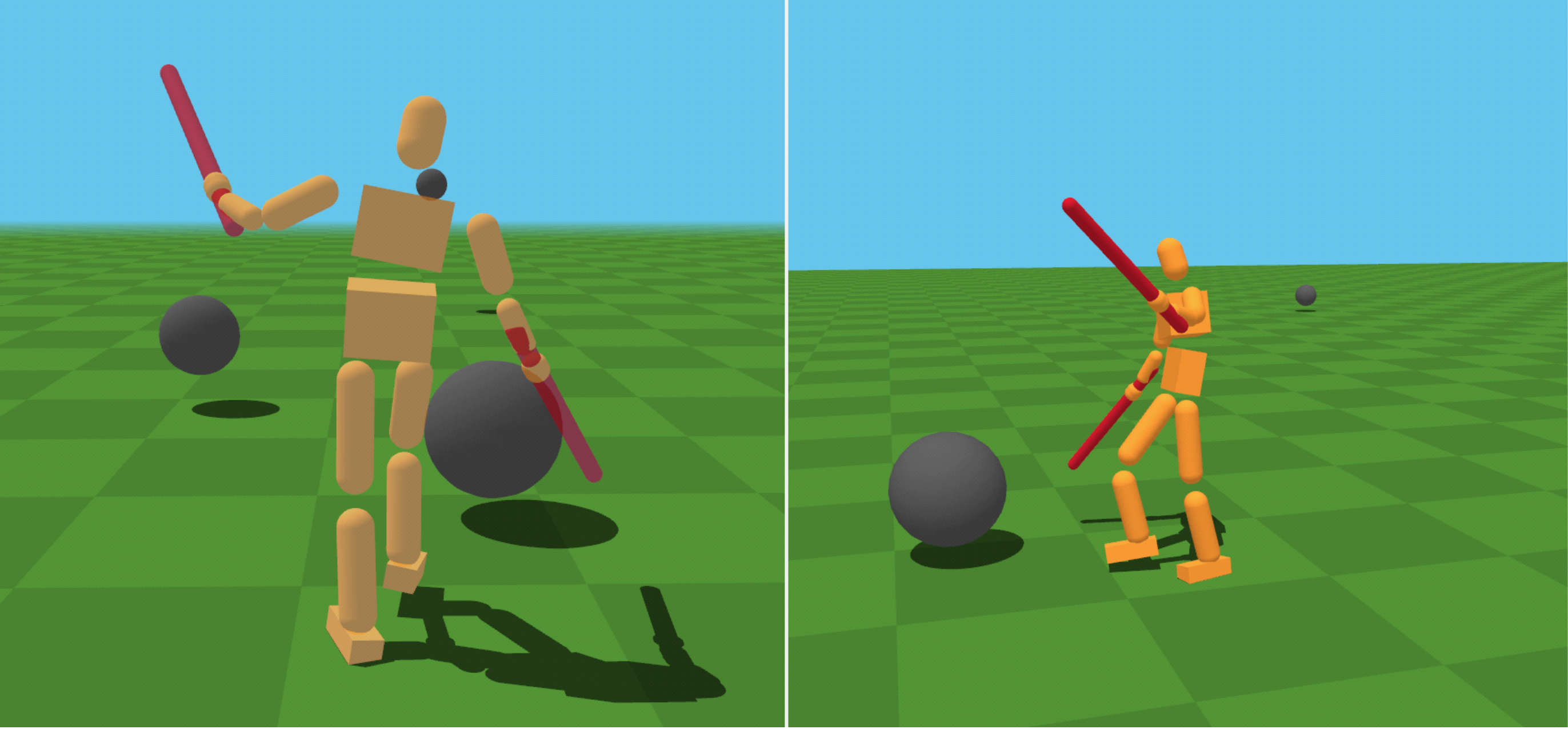}
   \caption{\label{fig:demo-game-1} Demo game \#1.
    A user need to block or avoid rolling stones to survive.
   }
\end{figure}

In the first game, a user needs to survive from the endless stones rolling at the game character (Figure~\ref{fig:demo-game-1}).
The user can avoid stones by leaning the body or blocking the stones using the two sticks in the character's hand.
Each stick is $80$ cm long and weighs $0.1$ kg.
The stones are randomly generated with a radius range of $[5, 30]$ cm, a mass range of $[0.02, 4]$ kg (which is proportional to their volume), and an initial speed range of $[5, 10]$ m/s.
Without any additional game elements such as other skills or items, the game demonstrates its own gameplay using performance-based control and a physics simulation.

\begin{figure}
   \centering
   \includegraphics[trim= 50 170 150 50, clip, width=1.\columnwidth]{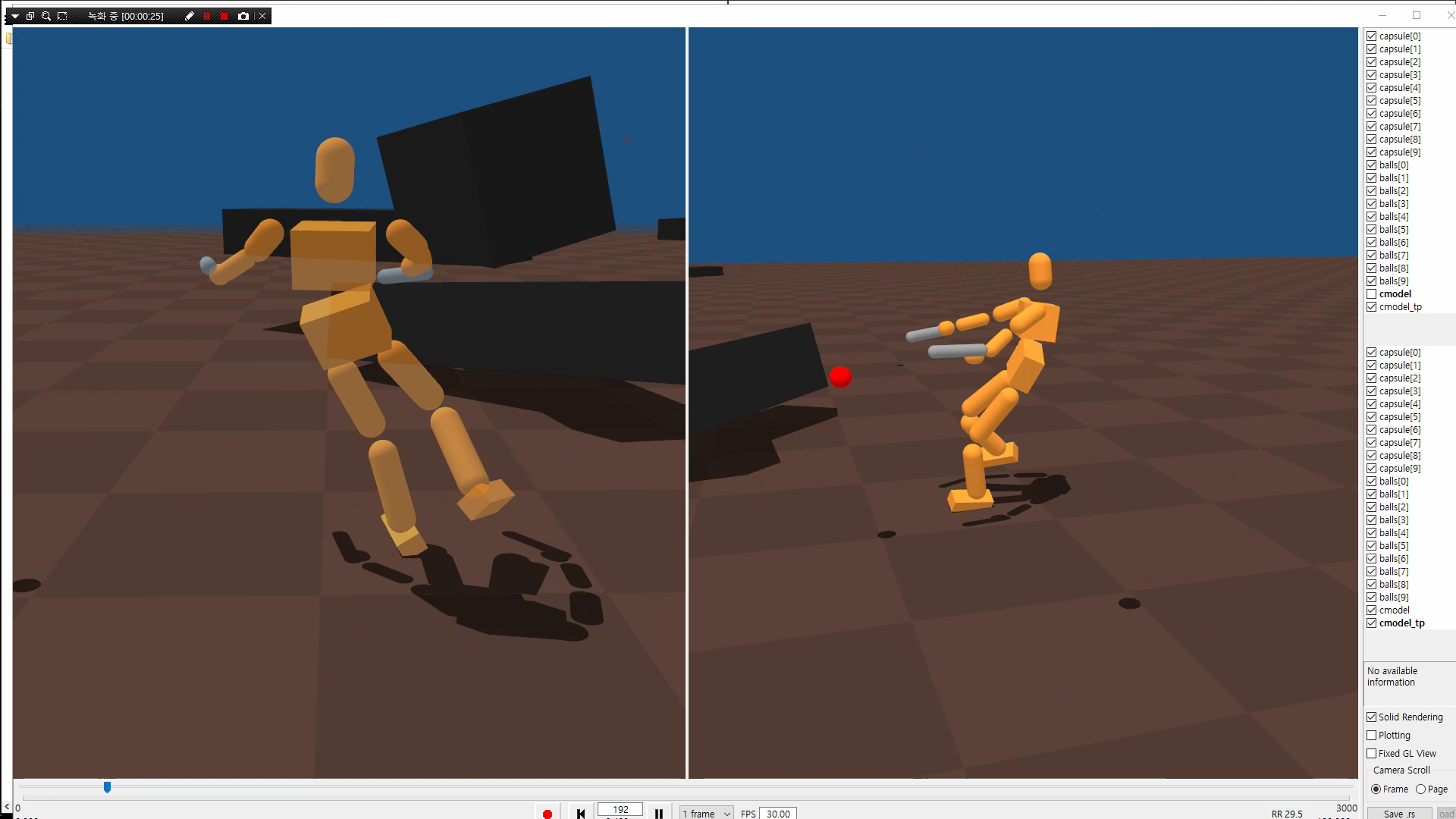}
   \caption{\label{fig:demo-game-2} Demo game \#2.
    A user has to use their guns to protect themselves from the boxes.
   }
\end{figure}

In the second game, the character encounters huge boxes randomly thrown toward it (Figure~\ref{fig:demo-game-2}).
The user can shoot guns to push away the boxes.
We use the Kinect's grasp detection to shoot the gun.
The speed of a bullet is $20$ m/s, and a box weighs between $1$ kg to $20$ kg randomly.
The games are controlled based on user performance and optionally, hand gestures, 
and are actually fun to play.

\begin{table}
	\centering
	\resizebox{1\columnwidth}{!}{
		\def\arraystretch{1.1}
		\setlength{\tabcolsep}{6pt}
		\begin{tabular}{|c|c|c|}
			\hline
            & \specialcell{Computation time \\(hours)} & \specialcell{Avg. \# of \\ CMA iteration} \bigstrut\\
			\hline
            \specialcell{Initial points    (24/45)} & 21.5 & 157.4  \bigstrut\\
			\hline
            \specialcell{Scaled by $0.8$   (16/21)} & 11.5 & 146.9  \bigstrut\\
			\hline
            \specialcell{Scaled by $0.8^2$ (4/5)} & 2 & 137.4  \bigstrut\\
			\hline
            \specialcell{Scaled by $0.8^3$ (0/1)} & 0.6 & 200  \bigstrut\\
			\hline
            \specialcell{Scaled by $0.8^4$ (1/1)} & 0.2 & 68  \bigstrut\\
			\hline
		\end{tabular}%
	}
	\caption{\label{tbl:comp-time}%
        Computation time for optimizing the parameter sample points.
        Starting from $45$ points, we repeatedly scaled down the failed points by $0.8$ until a successful solution is found.
        (a/b) denotes the number of successful points (a) and total points (b) at each stage.
    }
\end{table}

\textbf{Computation Time.} 
Because the entire optimization process includes a large number of time-consuming optimizations, we carefully determined the terminating criteria for CMA optimization to reduce the computation time.
The evaluated value of the objective function (Equation~\ref{eq:objective_func}) $2000$ \textcolor{rev}{was experimentally chosen to make walking simulation almost as stable as when the optimal objective value was used.
In most cases, stability improvement was negligible below that value despite consuming too much computation time and the value was found in $200$ iterations.}
Using these criteria, the entire computation process takes about $35$ hours using a single workstation.
The detailed time is given in Table~\ref{tbl:comp-time}.

\section{Discussion}

Using our approach, a user can freely swing his or her arms while controlling a walking biped.
Simply tracking the user pose from a depth camera would result in falls of the simulated biped.
Real-time interactive control is made possible by using a regression on the precomputed set of parameters and a balancing mechanism that compensates for the deviation between the simulated character and the user.

In the accompanying video, the biped exhibits some jerky upper-body motions due to the noise in the Kinect data.
We leave them as they are because handling input noise is not a goal of this paper, and the jerkiness does not affect the stability of our simulated biped.
It should also be noted that the skeleton data from the current generation depth camera is not suitable for synthesizing quality animations. 
There are a lot of problems such as noises, incorrect labels, unnatural rolls of limbs, and over-extended elbows and knees. 
Better performing input devices would improve the motion quality, and also encourage the development of other interesting real-world applications.

We assumed that the space for input motions can be parameterized using only a small number of quasi-static quantities such as the deviation of the CM position from the reference motion. 
Although this assumption was made due to limited computational resources, \textcolor{rev}{the proposed 3D parameter space is effective for controlling a consistently walking biped which mimics user actions.
However, our controller may not show good results for some cases when there are abrupt changes in the upper-body pose, as shown in the accompanying video.
Additional dimensions for the parameter space would help the controller deal with such cases.}
For example, the velocity and acceleration of the CM are important for balancing, and the external forces could also be parameterized to improve the foot-step planning for balance recovery.
However, the training data for such parameters can only be generated in the form of a short motion clip, and thus obtaining steady controllers would require time-varying controller parameters such as the approach proposed by Liu et al.~\shortcite{2016-TOG-controlGraphs}.
\textcolor{rev}{
Also, the increased search space would require more effective methods such as a better regression method that shows a good performance with a small number of samples, or an adaptive sampling of the parameter space.}
The hybrid use of pre-computed searches and online searches such as those in ~\cite{Mordatch:2010:SIGGRAPH,Hamalainen:2015:OCS} would also be an interesting future research topic.

\textcolor{rev}{
In our demonstrations, the biped only exhibits forward walking at the same speed.
We chose normal walking because we experimentally found it was the most stable one, and focused on designing the controller that mimics user performance while walking consistently.
Because the controller proposed by Lee et al.~\shortcite{Lee:2010} can control various styles of walking, turning, and spinning as shown in their original demonstration, our performance-based controller can be extended to control those skills in principle.
Walking speed can also be adjusted by kinematically editing the reference motions.
However, to support these skills, a user interface needs to be extended to control the additional parameters (turning angles, forward speed, etc.).
Designing such an interface using a Kinect sensor is non-trivial because of the limited capture space, the limited recognition range of forward-facing direction, and the increased dimensionality of the parameter space.
Dealing with these issues would be an interesting topic for future research.
}

We use the stepping strategy built on top of a single normal walking motion for balance recovery. However, it is well-known that a
human uses a combination of multiple strategies including hip, ankle and stepping strategies depending on the amplitude and duration of the
perturbation. Also, the strategy depends on the style of the motion that the user is performing. For example, when practicing a Kendo skill, 
the agility and speed rely highly on the coordinated use of
the arms holding the sword and the legs for stepping.
A trajectory library approach to automatically generating and controlling
lower-body motions based on the input upper-body motion would also be promising.

\textcolor{rev}{
\section*{Acknowledgements}
We thank the anonymous reviewers for helpful feedback.
This work was supported by Basic Science Research Program through the National Research Foundation of Korea (NRF) funded by the Ministry of Education (NRF-2016R1D1A1B03930746) and the MSIP (NRF-2014R1A1A1038386), and by the Research Grant of Kwangwoon University in 2016.
}

\bibliographystyle{eg-alpha}

\bibliography{kinectwalker}

\newpage

\end{document}